\documentclass{article}

\usepackage{arxiv}

\usepackage[utf8]{inputenc} 
\usepackage[T1]{fontenc}    
\usepackage{hyperref}       
\usepackage{url}            
\usepackage{booktabs}       
\usepackage{amsfonts}       
\usepackage{nicefrac}       
\usepackage{microtype}      
\usepackage{graphicx}
\usepackage{natbib}
\usepackage{doi}

\DeclareMathAlphabet\mathbfcal{OMS}{cmsy}{b}{n}
\usepackage{bm}
\usepackage[mathscr]{euscript}
\usepackage{enumitem}

\usepackage{hyperref}
\usepackage{amsmath}
\usepackage{amssymb}

\usepackage{amsthm}
\usepackage{stmaryrd}
\usepackage{dsfont}
\usepackage{xfrac}
\usepackage{mathrsfs}

\title{TriADA: Massively Parallel Trilinear Matrix-by-Tensor Multiply-Add Algorithm and Device Architecture for the Acceleration of 3D Discrete Transformations}

\date{} 					

\author{ \href{https://orcid.org/0000-0002-0071-5140}{\includegraphics[scale=0.06]{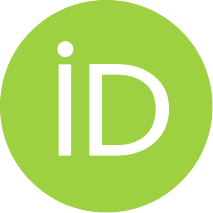}\hspace{1mm}Stanislav~Sedukhin}
\\
	The University of Aizu\\
    Fukushima, Japan\\
	\texttt{c21stans@u-aizu.ac.jp} \\
	\And
	\href{https://orcid.org/0000-0003-3509-6607}{\includegraphics[scale=0.06]{orcid.pdf}\hspace{1mm}Yoichi~Tomioka} 
\\
	The University of Aizu\\
    Fukushima, Japan\\
	\texttt{ytomioka@u-aizu.ac.jp} \\
	\AND
	 Kazuya Matsumoto \\
	 The University of Aizu \\
     Fukushima, Japan\\
	\texttt{kazuya-m@u-aizu.ac.jp} \\
	\And
	Yuichi Okuyama \\
	 The University of Aizu \\
	 Fukushima, Japan \\
	 \texttt{okuyama@u-aizu.ac.jp} \\
}

\renewcommand{\shorttitle}{TriADA: Algorithm and Device Architecture for Acceleration of 3D Transforms}

\hypersetup{
pdftitle={TriADA},
pdfsubject={q-bio.NC, q-bio.QM},
pdfauthor={Stanislav Sedukhin},
pdfkeywords={multilinear transforms, vector/streaming algorithms, massively parallel computing, application specific processors, cellular architectures, crossover mesh network, decoupled stream memory.},
}

\begin{document}
\maketitle

\begin{abstract}

Multilinear transformations are key in high-performance computing (HPC) and artificial intelligence (AI) workloads, where data is represented as tensors. However, their high computational and memory demands, which grow with dimensionality, often slow down critical tasks. Moreover, scaling computation by enlarging the number of parallel processing units substantially increases energy consumption, limiting widespread adoption, especially for sparse data, which is common in HPC and AI applications. This paper introduces the Trilinear Algorithm and isomorphic to algorithm Device Architecture (TriADA) to address these challenges with the following innovations: (1) a massively parallel, low-rank algorithm for computing a family of trilinear (3D) discrete  orthogonal transformations (3D-DXTs), which is a special case of the more general 3-mode matrix-by-tensor multiplication (3D-GEMT); (2) a new outer-product-based GEMM kernel with decoupled streaming active memory, specially designed to accelerate 3D-GEMT operation; (3) an isomorphic to the proposed algorithm, fully distributed 3D network of mesh interconnected processing elements or cells with a coordinate-free, data-driven local processing activity, which is independent of problem size; (4) an elastic sparse outer-product (ESOP) method that avoids unnecessary computing and communication operations with zero-valued operands, thereby enhancing energy efficiency, computational accuracy, and stability. TriADA is capable of performing a variety of trilinear transformations with hypercubic arithmetic complexity in a linear number of time-steps. Our algorithm-to-accelerator co-design method maps the three 4D index or iteration spaces of a trilinear algorithm to the same unified 3D processing/storage/communication network, supporting massively parallel computing and data movement/reuse within the TriADA system. The massively parallel, scalable, and energy-efficient architecture of TriADA is ideal for accelerating multilinear tensor operations, which are the most demanding parts of AI and HPC workloads. 

\end{abstract}


\section{Introduction}
Linear, bilinear, and trilinear discrete transformations (DT), collectively named here as multilinear transformations, are known to play a fundamental role in many application domains such as signal and image processing, data compression, medical diagnostics, astrophysics, molecular dynamics, deep learning, etc. 
Previously, increasing demands for high-speed execution of many real-world applications have stimulated the development of a number of Fast Transform (FT) algorithms, such as Fast Fourier Transform (FFT) and its variants, with a dramatic reduction in arithmetic complexity \cite{CooleyTukey, 814652}. These recursion-based FT-algorithms have been deeply serialized by restricting data reuse almost entirely to take advantage of the sequential processing. 
Recursion, however, does require the problem's size to be equal to power-of-two, which significantly limits the generality of fast algorithms.  Moreover, in the majority cases, implementation of the bilinear and trilinear transforms requires the initial matrix and tensor data, respectively, to be square to simplify the needed one or two data transpositions.  This additionally limits a generality of fast transforms. 

Multilinear transformations play a significant role in both high-performance computing (HPC) and artificial intelligence (AI) workloads. In these advanced domains, data are inherently represented by tensors, which are multidimensional coordinated arrays capable of capturing complex relationships and structures within vast datasets.  
The compute-intensive transformations of multidimensional data in AI workloads precede relatively simple element-wise nonlinear activation operations such as ReLU (sparse outputs), SquaredReLU (sparse outputs)~\cite{so2022primersearchingefficienttransformers}, sigmoid (dense outputs), tanh (dense outputs), GELU (dense outputs)~\cite{hendrycks2023gaussianerrorlinearunits} and many others~\cite{10.1016/j.neucom.2022.06.111}.
Linear and multilinear transformations of a vector, matrix, or tensor data are mathematically expressed in the form of matrix-by-vector, matrix-by-matrix, and matrix-by-tensor multiplications, respectively. The current demand for acceleration of these transformations in AI workloads is sufficiently surpassed by that of HPC workloads.
While HPC workloads involve mainly orthogonal or change-of-basis linear and multilinear transformations with invertible well-structured matrices of {\em predefined} coefficients, AI computing workloads involve mainly nonorthogonal and nonsquare matrices with {\em learnable} by very expensive training coefficients or parameters. This is why the fast and efficient matrix multiply-add units become paramount for hardware acceleration of today's training and inference parts of AI (super)computing.   
Another distinguishing feature of AI workloads is an extremely high degree of unstructured data sparsity, which can dramatically affect the computing and energy efficiency of data processing and storage.

The linear transformation parts in AI workloads, presented in the form of generalized matrix-by-matrix (GEMM) multiplications, are accelerated today by a large variety of hardware chips from leading industrial vendors such as AMD Matrix Cores~\cite{10590025}, Apple M-series with Advanced Matrix Extensions~\cite{hübner2025applevsorangesevaluating}, IBM Power~9 with Matrix Engines~\cite{9796646}, Google Tensor Processing Units~\cite{jouppi2023tpuv4opticallyreconfigurable}, Intel processors with Advanced Matrix Extensions~\cite{enwiki:1281191096}, NVIDIA GPUs with Tensor Cores~\cite{luo2024benchmarkingdissectingnvidiahopper} and many more~\cite{domke2021matrixengineshighperformance, pilz2025trendsaisupercomputers, electronics13152988, 9739030}. Unfortunately, these accelerators are designed for the fast and efficient execution of GEMM operation, that is, directly support only linear transforms of matrix data and do not well-suited for computing the bilinear and/or trilinear transforms of tensor data. 

The further speeding up of FT algorithms becomes possible by using an inherent concurrency of atomic operations with deep data reuse. However, it is well-known that due to serialization, FT algorithms have a very low degree of data reuse or the so-called arithmetic intensity, making these algorithms memory bounded. This obstacle prevents any efficient extremely parallel execution of FT algorithms on the computers and domain-specific accelerators with the number of arithmetic units (cores) equal to or close to the problem size, i.e., when each core stores and computes at each time-step only one or limited number of important for the algorithm resulting elements.  

In general, storing multidimensional data in main memory, with the linear address space, does destroy the multi-mode data integrity or locality, which has to be restored and used for massively parallel data computing, communication and reusing. The approach proposed in this paper trying to eliminate the above-mentioned limitations, first, by storing the elements of 3-mode data tensor in extremely distributed 3D grid of processing-storage-communication cells (one element per cell) and, second, by massively parallel data computing and reusing/movement within a 3D interconnected network of cells. The rectangular 3D shape of this network is isomorphic to the initial, intermediate, and final shapes of the computed data tensors. 

 Previously, we have proposed an algorithm-accelerator co-design methodology for extreme strong scaling trilinear discrete transformations on a 3D toroidal network of simple cores~\cite{sedukhin_co-design_2012, sedukhin_patent_2009}. The fused computational and local data moving kernels of these transformations have been based on different block or volumetric variants of Cannon-like algorithms for massively parallel and efficient multiplication of the {\it{square}} matrices~\cite{cannon_cellular_1969, Sedukhin2010}. A resulting scalable three-stage 3D-DXT algorithm has been ported and evaluated on FPGA (3D Discrete Cosine Transforms)~\cite{ikegaki_3d-dct_2011}, on IBM Blue Gene/Q (3D DFT)~\cite{sakai_3d_nodate, sedukhin_3d_2015}, on the supercomputer Avitohol (3D DFT)~\cite{lirkov_performance_2020}, and on JUWELS Cluster (3D DFT)~\cite{malapally_scalability_2023, MALAPALLY2024104945}.  
 Unfortunately, these 3D DFT implementations were tested in a weak scaling mode on the throughput-oriented supercomputers when the number of cores or nodes was much less than the size of tested problem and when the subproblem size per node should be big enough to tolerate with the numerous harmful latencies of internode MPI-based communications, pipelined arithmetic units, memory hierarchy, OS, etc. Therefore, it was not possible to demonstrate the advantages of our fine-grained approach with respect to the highly optimized course-grained 3D FFT libraries. At the same time, the execution run-time difference was already much less than the expected ideal DT/FT ratio of $\mathcal O(N/log N)$ for the relatively large number of interconnected nodes.
 
Additionally, real-life applications require the acceleration of computing of not only cubical but also mainly cuboid data tensors~\cite{di2025surveyerrorboundedlossycompression}. For example, for biomolecular simulations, the sizes of any of the three tensor dimensions are typically between 32 and 128 and may not be power-of-two~\cite{bowers_scalable_2006}.  In Deep Learning, which can use a 3D FFT~\cite{lin_fft-based_2018,fang_accelerating_2021}, the shape of the tensor changes from layer to layer of the Deep Neural Network~\cite{sedukhin_search_2022}. It is clear that Cannon-like algorithms cannot be used for efficient acceleration of a {\it{rectangular}} matrix or tensor data multiplication. 
Moreover, the previously proposed algorithm requires that three coefficient or change-of-basis matrices be (1) extended to cubical tensors by data replication and (2) distributed in advance among a 3D network of processing elements (PEs) or cells. This arrangement is based on the three-mode generalized tensor-by-tensor (GETT) multiplication which reuses at each time-step the two input tensors to update all elements of output tensor. Data reusing is implemented by a systolic-like cyclical shift or roll of {\bf two tensors} in each step. Therefore, to update all elements or points in the resulting {\em{cubical tensor}}, two, also cubical, input tensors should be {\em locally} moved inside a 3D toroidal processor@storage network. This collective shift of two data tensors on each time-step of 3D communication introduces a certain overhead, which can be considered as the algorithm's drawback.

In this paper, we propose a new outer-product-based or low-rank algorithm that supports efficient, extremely parallel three-mode or three-dimensional generalized matrix-by-tensor (3D-GEMT) multiplication on a fully distributed 3D network of mesh interconnected cells. To update all elements in the output {\em{rectangular tensor}}, this new algorithm requires in each time-step the reusing by {\em{global}} replication of only a single {\bf vector and matrix}. 
We will use our previously proposed algorithm-accelerator co-design methodology for the same Fourier-like family of trilinear orthogonal transformations~\cite{sedukhin_co-design_2012}. In general, all these transformations require multiplication of three square coefficient matrices by non-square or rectangular tensor, which is a special case of the more general 3D-GEMT algorithm, where matrices might also be non-square.

The proposed Trilinear Algorithm and isomorphic Device Architecture (TriADA) can store and accelerate a variety of 3D discrete orthogonal transformations with hypercubical arithmetic complexity in a linear number of time-steps. 
TriADA integrates novel algorithmic advances with a specialized, distributed hardware architecture, aiming to fundamentally mitigate computational, communicational, memory, and energy constraints.
This architecture is structurally aligned or morphed with the proposed algorithm, enabling a coordinate-free and problem-size-independent computation. The term "isomorphic" is critical here, signifying that the hardware architecture is structurally designed to mirror the algorithm's operational flow and data dependences.

The algorithm-architecture co-design involves mapping three 4D index or coordinate spaces
of the 3D-DXT/GEMT algorithm onto the same unified 3D processing-storage space of TriADA. This enables highly parallel
computing and data movement/reuse entirely within the TriADA network. The proposed low-rank implementation of the 3D-DXT
algorithm and the isomorphic TriADA network are well-suited for accelerating multilinear tensor operations, which constitute the
most computationally intensive parts of today's AI and HPC workloads. Furthermore, the algorithm and accelerator architecture are
simple, regular, homogeneous, scalable, and exhibit high data reusability, positioning them well for wafer-scale computing~\cite{10460211, he2025waferllmwaferscalellminference, swartzlander1989wafer, 10.1145/3577193.3593708} and the upcoming trillion-transistor era~\cite{10589682}.

\section{Trilinear Orthogonal Transforms}
\subsection{Basic Equations}
Let $\mathbfcal X=\llbracket x_{n_1,n_2,n_3}\rrbracket$, $n_1\in [0,N_1)$,  $n_2\in [0,N_2)$,  $n_3\in [0,N_3)$, 
be an $N_1{\times} N_2{\times} N_3$ Cartesian grid of input data elements or a three-dimensional (3D) data tensor~\cite{kolda_tensor_2009}.
A separable {\it forward} 3D linear transformation changes a tensor coordinate system $\mathbfcal X$ to the new coordinates $k_1\in [0, N_1)$, $k_2\in [0, N_2)$, $k_3\in [0, N_3)$, by computing another grid of a $N_1{\times} N_2{\times} N_3$ data or three-mode tensor $\mathbfcal{\dddot{X}}=\llbracket\dddot{x}_{k_1,k_2,k_3}\rrbracket,$ where each single data element $\dddot{x}_{k_{1},k_{2},k_{3}} $ is computed in the {\em generalized} multiply-add form as 
\begin{eqnarray}
  \dddot{x}_{k_{1},k_{2},k_{3}} \mathrel{{+}{=}} \sum_{n_{3}=0}^{N_3-1} \sum_{n_{2}=0}^{N_2-1} \sum_{n_{1}=0}^{N_1-1} x_{n_{1},n_{2},n_{3}} 
  c_{n_{1},k_{1}}  c_{n_{2},k_{2}}  c_{n_{3},k_{3}}~.
  \label{eq:separable_forward_3d} 
\end{eqnarray}
Here, elements of the output tensor $\mathbfcal{\dddot{X}}$ should be initialized at the beginning of processing, and, in general,  $\mathbfcal{\dddot{X}}$ might initially not be a zero tensor. This required initialization extends a linear transformation to the more general {\em affine} transformation by using $(\mathrel{{+}{=}})$ notation in Eq.~(\ref{eq:separable_forward_3d}) which combines the linear maps (=) and translations or shifts (+). 

In turn, a separable {\em inverse} or {\it backward} 3D transformation of the three-dimensional input tensor $\mathbfcal{\dddot{X}}=\llbracket\dddot{x}_{k_1,k_2,k_3}\rrbracket$ is expressed for each output component $x_{n_{1},n_{2},n_{3}}$ as 
\begin{eqnarray}
  x_{n_{1},n_{2},n_{3}} \mathrel{{+}{=}} \sum_{k_{3}=0}^{N_3-1} \sum_{k_{2}=0}^{N_2-1} \sum_{k_{1}=0}^{N_1-1} \dddot{x}_{k_{1},k_{2},k_{3}} 
  c_{n_{1},k_{1}}  c_{n_{2},k_{2}}  c_{n_{3},k_{3}}~,
  \label{eq:separable_inverse_3d}
\end{eqnarray} 
where $n_1\in [0, N_1)$, $n_2\in [0,  N_2)$, $n_3\in [0, N_3)$, and $\mathbfcal X = \llbracket x_{n_1,n_2,n_3}\rrbracket$ is a reconstructed $N_1{\times}N_2{\times}N_3$ tensor. 

In Eqs.~(\ref{eq:separable_forward_3d}) and (\ref{eq:separable_inverse_3d}), $[c_{n_1,k_1}]={\bf C}_{N_1\times N_1}$, $[c_{n_2,k_2}]={\bf C}_{N_2\times N_2}$, $[c_{n_3,k_3}]={\bf C}_{N_3\times N_3}$ are the square, orthogonal, invertible coefficients or change-of-basis matrices.
The calculation of (\ref{eq:separable_forward_3d}) or (\ref{eq:separable_inverse_3d}) is based on three summations or reductions along the corresponding tensor dimensions or modes. Note that (\ref{eq:separable_forward_3d}) and (\ref{eq:separable_inverse_3d}) are the element-wise notations of computing.

\subsection{Types and Arithmetic Complexity of Trilinear Orthogonal Transformations}
\label{types}
The various separable orthogonal transformations differ only by the coefficient or change-of-basis, square and invertible, $(N{\times}N)$-matrix ${\bf C}=[c(n,k)]$, which can be either
\begin{itemize}
  \begin{item}
    {\it symmetric}, i.e. ${\bf C}={\bf C}^{\top}$, and {\it unitary}, i.e. ${\bf C}^{-1} = {\bf C}^{*\top}$, ${\bf C}^{*}$ is a complex conjugate of ${\bf C}$, like in the Discrete Fourier Transform (DFT), where $c_{n,k} = \exp[-\frac{2\pi i}{N}(n k)] = \cos(\frac{2\pi n k}{N})-i\sin(\frac{2\pi n k}{N})$ and $i = \sqrt[]{-1}$, or in the Discrete Hartley Transform (DHT), where $c_{n,k} = \cos(\frac{2\pi n k}{N})+\sin(\frac{2\pi n k}{N})$; 
  \end{item}
  \begin{item}
    unitary and {\it real}, i.e. {\it orthogonal}, like in the Discrete Cosine Transform (DCT), where coefficient $c_{n,k} = cos[\frac{\pi}{2N}(2n+1) k]$ and ${\bf C} \neq {\bf C}^{\top}$; 
  \end{item}
  \begin{item}
    consists only $\pm$ 1 and be symmetric and orthogonal, like in the Discrete Walsh-Hadamard Transform (DWHT). 
  \end{item}
\end{itemize}
The elements of these coefficient matrices can be pre-computed, stored and reused as constants or computed in the required time~\cite{289998}. 
As can be seen, only the very popular Fourier transform requires complex numbers and complex arithmetic, whereas all other inherently similar transformations have been developed in order to avoid these complex data and operations. Note that coefficients of these matrices (its entries) can include zeros as long as the orthonormality condition is met.
We will abbreviate this common for these 3D discrete transformations computational framework as 3D DXT.

Direct element-wise computing of Eq.~(\ref{eq:separable_forward_3d}) or Eq.~(\ref{eq:separable_inverse_3d}), with a 6-dimensional (6D) index or coordinate space each, requires execution of $(N_1N_2N_3)^2$ ``multiply-add combined'' (MAC) operations which are equal to the number of index points in a 6D coordinate space.  Element-wise computing is usually expressed in a 6-nested loop program with an innermost MAC operation. Note that multilinear discrete orthogonal transforms can make data sparse, meaning most data elements in the transformed domain are zero, which helps with compression and faster computation. 

\subsection{A Three-mode GEMT Multiplication} 

There is direct correspondence between a trilinear orthogonal transformation Eq.~(\ref{eq:separable_forward_3d}) or Eq.~(\ref{eq:separable_inverse_3d}) and a {\it{three-mode matrix-by-tensor multiplication}}~\cite{kolda_tensor_2009,lim_tensors_2021}, which is used to represent the three-dimensional data tensor $\mathbfcal X$ or $\mathbfcal{\dddot X}$ in different bases and where each square matrix ${\bf C}_{N_s\times N_s}, s \in \{1, 2, 3\}$, is a non-singular orthogonal or change-of-basis matrix of coefficients. Recall that orthogonal transformations are distance-preserving or isometrical linear transformations which correspond to square matrices $\bf C$ that satisfy $\bf{C}^{\top}\bf{C}=I$, where $\bf{C}^{\top}$ is the transpose of $\bf C$ and $\bf I$ is the identity matrix. Such matrices represent rotations, reflections, or combinations thereof, and they preserve lengths and angles. This distance conservation is tightly connected to preserving the inner product, which preserves norms and thus distances. 

However, most linear transformations, such as stretching, shearing, or projections, do not preserve distances. For example, in deep learning, both linear and multilinear transformations are generally not orthogonal, as weight or coefficient matrices and tensors are optimized without constraints on preserving distances or angles.
Therefore, in the more general 3-mode matrix-by-tensor (3D-GEMT) multiplication, the coefficient matrices can be non-square and not orthogonal, but rectangular matrices ${\bf C}_{N_s{\times}K_s}$ with any number of columns $K_s$. Like in 3D DXT, in trilinear matrix-by-tensor multiplication, each of the three matrices is multiplied by a 3D tensor along one of its modes (dimensions). 

The execution of a 3D-GEMT operation also includes tensor expansion,
when $K_s>N_s$, and tensor compression, when $K_s<N_s$, $s\in[1,2,3]$. The last inequality relates to Tucker's tensor decomposition ~\cite{kolda_tensor_2009, lim_tensors_2021}, where the initial $(N_1{\times}N_2{\times}N_3)$-tensor is approximated  in the form of multiplication of the so-called core $(K_1{\times}K_2{\times}K_3)$-tensor by three rectangular coefficient or factor matrices
${\bf C}_{K_1{\times}N_1}$, ${\bf C}_{K_2{\times}N_2}$, and ${\bf C}_{K_3{\times}N_3}$. Finding the appropriate factor matrices with $K_s\ll N_s$, $s=1,2,3$, is a very important and challenging problem in multilinear algebra~\cite{de_lathauwer_multilinear_2000, 7891546,RUTLEDGE2007170}.
\href{https://en.wikipedia.org/wiki/Tucker_decomposition}{Tucker decomposition} is frequently used for tensor contraction or model compression in abinitio quantum chemistry models \cite{1386652, LU2012338} as well as 
in modern machine learning workloads \cite{Acar2009,10190238}. The proposed trilinear algorithm and accelerator architecture can also be used to compute the extension and compression of three-mode tensors.

\section{Tensor Partition and Generalized  Matrix-by-Tensor Multiplication} \label{tp}
\begin{figure}[htbp]
\begin{center}
\includegraphics[width=1\textwidth]{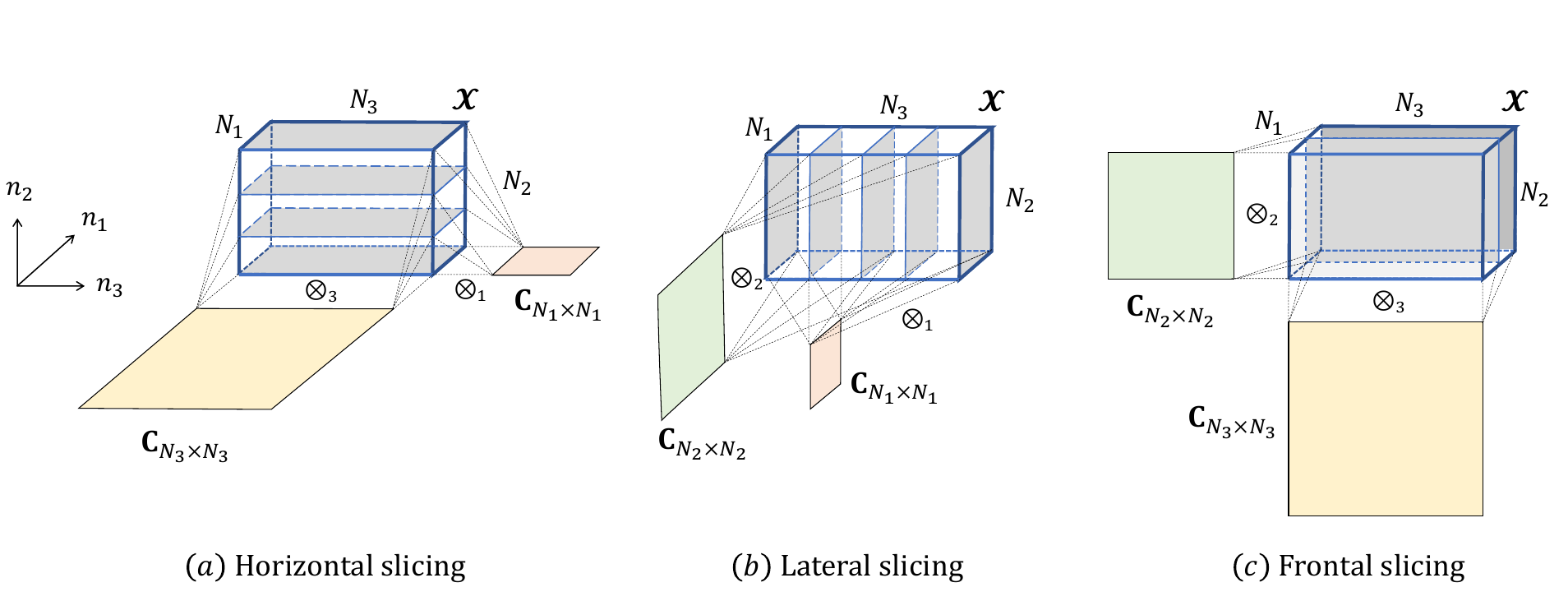}
\caption{Partitions of a 3-mode tensor into the set of slices or matrices (shown in gray) and corresponding two-mode GEMT multiplication where coefficient matrices are shared between all slices in this partition.}
\label{partition}
\end{center}
\end{figure}

A three-mode or 3D tensor can be partitioned into different non-overlapped sets of planar slices or matrices, which can be used to perform multiple independent linear or bilinear transformations.  
Figure~\ref{partition} shows the horizontal or $n_2$-direction slicing $(a)$, where $ \mathbfcal X=\bigcup_{n_2=0}^{N_2-1}{\bf X}_{N_1{\times}N_3}^{(n_2)}$,  the lateral or $n_3$-direction slicing $(b)$, where  $\mathbfcal X=\bigcup_{n_3=0}^{N_3-1}{\bf X}_{N_1{\times}N_2}^{(n_3)}$, and  the frontal  or $n_1$-direction slicing $(c)$, where $\mathbfcal X=\bigcup_{n_1=0}^{N_1-1}{\bf X}_{N_2{\times}N_3}^{(n_1)}$. 

Computing the trilinear or three-mode GEMT multiplication (\ref{eq:separable_forward_3d}) or (\ref{eq:separable_inverse_3d}) by using these partitions and the corresponding numbers of the sliced matrix-by-matrix (GEMM)  multiplications\footnote{Recall that each matrix-by-matrix multiplication has a 3D index space.} reduces the dimensionality of the iteration space from the monolithic 6D space to three, partially overlapped by computing the intermediate data, 4D index spaces (see details in \cite{sedukhin_co-design_2012}). This reduction in dimensionality decreases the number of required atomic MAC operations or, equally, the number of index points in a coordinate space, from $(N_1N_2N_3)^2$ to $N_1N_2N_3 (N_1{+}N_2{+}N_3)$ at the price of an additional memory space to store intermediate data tensors.  Partition or slicing of a tensor is closely related to the parenthesization of a 3-mode GEMT computing 
\begin{equation}
{\mathbfcal{\dddot X} }=\mathcal C_1^{\top}{\mathbfcal X}{\mathcal {C}_3\mathcal{C}_2},
\end{equation}
where $\mathbfcal X$, $\mathbfcal{\dddot X}$ are the input and output tensors of size $(N_1{\times}N_2{\times}N_3)$, respectively, and $\mathcal C_s={\bf C}_{N_s{\times}N_s}, s \in \{1,2,3\}$ are the coefficient matrices. Using, initially, a horizontal, a lateral, or a frontal tensor slicing, shown in Figures~\ref{partition} $(a), (b)$, and $(c)$, respectively, allows two possible computing parentheses each, i.e., totally, 6 permutations:
\begin{itemize}
\item a horizontal slicing first and then a lateral or a frontal slicing
	\begin{itemize}
	\item ${\mathbfcal{\dddot X} }=(\mathcal C_1^{\top}({\mathbfcal X}{\mathcal {C}_3))\mathcal{C}_2};$
	\item ${\mathbfcal{\dddot X} }=((\mathcal C_1^{\top}{\mathbfcal X}){\mathcal {C}_3)\mathcal{C}_2};$
	\end{itemize}
\item a lateral slicing first and then a horizontal or a frontal slicing 
	\begin{itemize}
	\item ${\mathbfcal{\dddot X} }=((\mathcal C_1^{\top}{\mathbfcal X}){\mathcal {C}_2)\mathcal{C}_3};$
	\item ${\mathbfcal{\dddot X} }=(\mathcal C_1^{\top}({\mathbfcal X}{\mathcal {C}_2))\mathcal{C}_3};$
	\end{itemize}
\item a frontal slicing first and then a horizontal or a lateral slicing
	\begin{itemize}
	\item ${\mathbfcal{\dddot X} }=\mathcal C_1^{\top}(({\mathbfcal X}{\mathcal {C}_2)\mathcal{C}_3)};$
	\item ${\mathbfcal{\dddot X} }=\mathcal C_1^{\top}(({\mathbfcal X}{\mathcal {C}_3))\mathcal{C}_2}).$
	\end{itemize}
\end{itemize}

In other words, the selection of the initial partition of a tensor defines the order of three required summations. Each planar slice or matrix in a tensor partition allows for the implementation of two out of three summations in any order. However, the remaining third summation requires the switching to a different tensor partition with missing direction of summation.

It is very important to mention that a GEMT notation of the trilinear transformation does not require the problem sizes $N_1$, $N_2$, $N_3$ to be either power-of-two, like in FFT, or prime numbers, or, most notable, be equal to each other. The existence of these limitations sufficiently reduces the generality of many so-called {\it fast} algorithms.

\subsection{Inner-Product Notation of a Three-mode GEMT Multiplication}

The shown below three-mode GEMT-based processing is based on the selection of the horizontal input tensor slicing, first, illustrated in  Figure~\ref{partition}({\it a}) for Stages~I and II  (with summations along the leading dimensions $n_3 \in [0,N_3)$ and $n_1 \in [0,N_1)$, respectively) and second, on the frontal reslicing of an intermediate tensor, shown in  Figure~\ref{partition}({\it c}) for Stage~III (with summation along the complimentary dimension $n_2\in[0,N_2)$):  

\begin{subequations}
\renewcommand{\theequation}{\theparentequation.\arabic{equation}}
\label{dp}
\begin{equation}
{\text{Stage~I.}}~\forall n_2\in[0,N_2): 
\Bigl\{\forall n_1\in[0,N_1), \forall k_3\in[0,N_3): {\dot x}_{n_1,k_3}^{(n_2)}\mathrel{{+}{=}}\sum_{n_3=0}^{N_3-1}x_{n_1,n_3}^{(n_2)}c_{n_3,k_3} 
\mathrel{{+}{=}} {\bf x}(n_1)_{N_3}^{(n_2)}\cdot {\bf c}(k_3)_{N_3}\Bigl\};
\label{dp1}
\end{equation}

\begin{equation}
{\text{Stage II.}}~\forall n_2\in[0,N_2):
\Bigl\{\forall k_1\in[0,N_1), \forall k_3\in[0,N_3): {\ddot x}_{k_1,k_3}^{(n_2)}\mathrel{{+}{=}}\sum_{n_1=0}^{N_1-1}c_{k_1,n_1}\dot x_{n_1,k_3}^{(n_2)} 
\mathrel{{+}{=}} {\bf c}(k_1)^{\top}_{N_1}\cdot \dot{\mathbf{x}} (k_3)_{N_1}^{(n_2)}\Bigr\};
\label{dp2}
\end{equation}

\begin{equation}
{\text{Stage III.}}~\forall k_3\in[0,N_3): 
\Bigl\{\forall k_1\in[0,N_1), \forall k_2\in[0,N_2): \dddot x_{k_1,k_2}^{(k_3)}\mathrel{{+}{=}}\sum_{n_2=0}^{N_2-1}\ddot x_{k_1,n_2}^{(k_3)} c_{n_2,k_2} 
 \mathrel{{+}{=}}\ddot{\mathbf{x}}(k_1)_{N_2}^{(k_3)}\cdot {\bf c}(k_2)_{N_2}\Bigr\}.
\label{dp3}
\end{equation}
\end{subequations}

In the equations above, each stage is presented not only in the traditional form as the sum of element-wise  products, but also in a vector, row-by-column, inner- or dot-product form $(\cdot)$. The lower index of the vector indicates the vector length or norm, while the upper index shows the slice's number.  This inner-product notation of computing requires both vectors to have equal lengths. Moreover, as can be seen from Eqs.~(\ref{dp}), each coefficient matrix is shared among all tensor slices of the given stage.    

Note that to correctly implement matrix-by-matrix multiplication in Stage~II, a coefficient matrix ${\bf C}_{N_1{\times}N_1}=[c_{n_1,k_1}]$ should be transposed and used in the left or multiplicand place, as can also be seen in  Eq.~(\ref{dp2}), where $c_{k_1,n_1}$ and ${\bf c}(k_1)_{N_1}$ are the single element and $k_1$-th column of a transposed matrix ${\bf C}^{\top}_{N_1{\times}N_1}$, respectively.
Furthermore, the implementation of Stage~III, which demands summation along dimension $n_2\in [0, N_2)$, requires a repartition of calculated in Stage~II intermediate tensor $\mathbfcal{\ddot{X}}=\llbracket\ddot{x}_{k_1,n_2,k_3}\rrbracket$. This repartition can be performed using a lateral or frontal slicing, shown in Figures~\ref{partition}({\it b}) and \ref{partition}({\it c}), respectively.
As an example, a frontal slicing is used in Eq.~(\ref{dp3}), based on  the equality: 
\begin{equation}
\ddot{\mathbf{X}}_{N_1{\times}N_3}^{(n_2)} \ni {\ddot x}_{k_1,k_3}^{(n_2)} \equiv {\ddot x}_{k_1,n_2}^{(k_3)} \in \ddot{\mathbf{X}}_{N_1{\times}N_2}^{(k_3)}.
\label{repartition}
\end{equation}

It is clear that the first part of a processing chain, Eq.~(\ref{dp1}), defines computation of the {\em linear} transformation of a 3D data tensor. The second part of a chain, Eq.~(\ref{dp2}), relates to the {\em bilinear} tensor transformation, and the third, final part, Eq.~(\ref{dp3}), specifies computation of the {\em trilinear} tensor transformation.

\subsection{GEMM: Scalar, Vector and Matrix Notations }

It is known that the generalized matrix-by-matrix (GEMM) multiply-add operation can be formulated in scalar, vector, or matrix notation~\cite{golub_matrix_2013}.
\begin{enumerate}
\item 
A scalar or inner-product (IP) notation defines a direct and independent computation of each {\it scalar} result of the output matrix as the inner-product of two input vectors (row and column) of equal size as shown in Eqs.~(\ref{dp}). Here, the vector size is equal to the length of the summation. 
The total number of such data independent IP-operations is equal to the size of output matrix, i.e., quadratic.

\item A vector or so-called {\texttt{SAXPY}}\footnote{\underline{S}calar \underline A times vector \underline X \underline Plus vector \underline Y.} notation of a GEMM operation, is based on a scalar-by-vector product (SVP) or, in general,  scalar-by-vector multiply-add operation which is used multiple  times (equal to the vector length)  to update  a resulting row or column {\it vector} of output matrix.  Because all vectors of this output matrix should be computed, the total number of such SVP operations will be equal to the size of output matrix, i.e., again quadratic. Note that this variant, also known as Gustavson\textquotesingle{}s row-wise algorithm~\cite{10.1145/355791.355796}, is popular for computing a sparse GEMM by avoiding updating the output vector if a scalar equals zero \cite{10.1145/3445814.3446702, 10139817}.

\item A matrix or outer-product (OP) GEMM notation involves two (column-and-row) input vectors of different lengths, in general, to update all elements of an intermediate output {\it matrix}. In this case, to compute a resulting output matrix as the sum of all intermediate matrices,
 only the linear number of such OP or rank-1 update vector operations is needed.

\end{enumerate}

The three notations of the GEMM operation, namely {\em element}-wise, {\em vector}-wise and {\em matrix}-wise, require the same cubical number of the atomic scalar multiply-add combined (MAC) or fused multiply-add (FMA) operations. 
At the same time, each notation has distinct characteristics concerning aggregation of these atomic operations, data access and reuse patterns, and degree of operation concurrency.
If each IP-, SVP- or OP-vector operation is assumed to be executed in one time-step, i.e., independent of the length of involved vectors,  then the only OP-variant computes a resulting matrix in the linear number of time-steps. 
A fruitful, but mostly physically and technology specific, discussion about how realistic this assumption is out of the scope of this paper (see, however, related references with a proper discussion~\cite{1674948,6158974, pdpu23,10323219,8758338}).

Also, based on a deep investigation by Google's AI assistant Gemini~\cite{gemini_gemm_notation_report}, the {\em outer-product} notation emerges as the most promising and useful for accelerating GEMM on both current and future hardware architectures. It offers better inherent support for tiling, excellent parallelism on GPUs, and a direct and natural mapping to the operations performed by tensor/matrix cores, which are increasingly becoming fundamental building blocks of high-performance computing, especially in the domains of artificial intelligence and machine learning. 
Moreover, in the scenario where there is a large number of simple computational cores, each potentially with a small integrated memory, and the goal is to perform the GEMM operation without the traditional need for tiling due to core scarcity, the outer-product formulation of GEMM also becomes the most promising. In addition, for the inner-product operation of two sparse vectors, it is challenging to avoid partial multiplications and additions of zero-valued randomly scattered vector operands. 

In the following, a new OP-formulation of the three-mode GEMT operation, adapted for computing 3D DXT, is presented and elaborated.


\subsection{Outer-product Notation of a Three-mode GEMT Multiplication}
A three-mode GEMT operation, expressed previously in a inner-product or element-wise notation~(\ref{dp}), can also be represented equally in a matrix-wise notation, that is, as the sum of the outer-products or rank-1 updates:
\begin{subequations}
\renewcommand{\theequation}{\theparentequation.\arabic{equation}}
\label{op}
\begin{equation}
{\text{Stage I.}}~ \forall{ n_2\in[0,N_2)}: {\dot {\bf X}}_{N_1{\times} N_3}^{(n_2)}\mathrel{{+}{=}}\sum_{n_3=0}^{N_3-1}{\bf x}(n_3)_{N_1}^{(n_2)}~\circ~{\bf c}(n_3)_{N_3};
\label{op1}
\end{equation}
\begin{equation}
\label{op2}
{\text{Stage II.}}~ \forall{n_2\in[0,N_2)}: {\ddot {\bf X}}_{N_1{\times}N_3}^{(n_2)}\mathrel{{+}{=}}\sum_{n_1=0}^{N_1-1}{\bf c}(n_1)_{N_1}~\circ~\dot{\bf x}(n_1)_{N_3}^{(n_2)};
\end{equation}
\begin{equation}
\label{op3}
{\text{Stage III.}}~\forall{k_3\in[0,N_3)}: \dddot {\bf X}_{N_1{\times} N_2}^{(k_3)}\mathrel{{+}{=}}\sum_{n_2=0}^{N_2-1}\ddot {\bf x}(n_2)_{N_1}^{(k_3)}~\circ~{\bf c}(n_2)_{N_2};
\end{equation}
\end{subequations}
where ${\bf x}(n_3)_{N_1}^{(n_2)}\in {\bf X}_{N_1{\times}N_3}^{(n_2)}$, $\dot{\bf x}(n_1)_{N_3}^{(n_2)}\in \dot {\bf X}_{N_1{\times}N_3}^{(n_2)}$, 
$\ddot {\bf x}(n_2)_{N_1}^{(k_3)}\in \ddot {\bf X}_{N_1{\times} N_2}^{(k_3)}$, and ${\bf c}(n_3)_{N_3}\in {\bf C}_{N_3{\times} N_3}$, ${\bf c}(n_1)_{N_1}\in {\bf C}^\top_{N_1{\times}N_1}$,
${\bf c}(n_2)_{N_2}\in {\bf C}_{N_2{\times}N_2}$ are {\it vectors} of the corresponding input, intermediate and final coefficient matrices that are involved in the related sum of  outer-products $(\circ)$.  
An outer-product notation assumes that, on each step of summation, the rectangular amount of atomic MAC operations is executed. This is why this notation is more compact than the inner-product one.

It can be seen by the lower index of the vectors in Eqs.~(\ref{op}) that now the lengths of these vectors are different in each stage. The intermediate and final matrices ${\dot {\bf X}}_{N_1{\times} N_3}^{(n_2)}$, ${\ddot {\bf X}}_{N_1{\times}N_3}^{(n_2)}$, and $\dddot {\bf X}_{N_1{\times} N_2}^{(k_3)}$ are calculated as {\it rank}--$N_3$, {\it rank}--$N_1$, and {\it rank}--$N_2$ {\it updates}, respectively.
Like in IP-notation, each coefficient matrix is shared among all tensor slices of a given stage. Recall that the outer-product notation of GEMM operation involves on each summation step a column-vector of the multiplicand (first) matrix and a row-vector of the multiplier (second) matrix. Note also that, unlike the inner-product, the outer-product is not commutative.

The dot-product notation, Eqs.~(\ref{dp}), and outer-product notation, Eq~(\ref{op}), of a 3D DXT computing as a 3-mode multiplication of square matrices by a non-square tensor, cover also the more general case of multiplication of non-square matrices by a non-square tensor (GEMT).

\section{Mapping Algorithm's Index Space to Processing@Storage Space}

As was shown in Section~\ref{tp}, the coordinate space of a 3D DXT algorithm consists of three overlapped 4D index spaces which cannot be directly mapped to the 3D physical implementation space.  
We use our previous approach for the reduction of the dimensionality of the algorithm index space by using a linear mapping of the $m$-dimensional spatial index or iteration space of an algorithm to the $(m-1)$-dimensional processor or implementation space~\cite{10.1007/3-540-58430-7_16, sedukhin_co-design_2012}.  Before this mapping, a time-scheduling function is defined, which assigns each index point from an iteration space the time-step of activation or computation. This function is selected such that it linearly depends on the point's indices or coordinates and does not conflict with the algorithm's data dependency. Indirectly, the time-step function also defines a conflict-free data movement between index points. This linear mapping is not unique. For example, the 3D iteration space of a matrix-by-matrix multiplication can be mapped into a variety of 2D array processors with different numbers of cells or processing elements (PEs), interconnection network, data flows, pipeline periods, etc.

Previously, a few time-scheduling functions have been formally introduced which plan both the spatial index-based computing and matrix data movement in many well-known planar array processors for acceleration of a dense GEMM operation~\cite{Sedukhin2010}:

\begin{enumerate}[label=\alph*)]
\item "compute-shift-all" or linear systolic-like schedule~\cite{kung_why_1982, kung__systolic_1985};
\item "compute-roll-all" or modular Cannon-like  schedule~\cite{cannon_cellular_1969};
\item "broadcast-compute-roll" or semi-modular Fox-like schedule~\cite{fox_matrix_1987};
\item "broadcast-broadcast-compute" or rank-1 update schedule~\cite{agarwal_high-performance_1994,van_de_geijn_summa:_1997,li_poly_1997, goto_anatomy_2008}.
\end{enumerate}

To spatially update elements of the output matrix in one time-step, these scheduling functions involve data reusing by movement either two input matrices in a) and b) cases or one input vector and one input matrix in c) case or two input vectors in d) case. 
Moreover, due to the modular ``roll" scheduling, the {b}) and {c}) functions can only be used for square matrices and/or tensors in 2D or 3D torus networks~\cite{289998, Sedukhin2010, sedukhin_co-design_2012}. 
Furthermore, while shift- and roll-based scheduling functions, in the {a}) and {b}) cases, respectively, define a local or short communication between nearest neighboring PEs, the {c}) and {d}) functions require global or long-distance data exchange/reuse via broadcast or multicast.
In this paper, the scheduling {d}), which is inherent in the outer-product computing of Eqs.~(\ref{op}), is used.

The spatial mapping of three rectangular 4D index or coordinate spaces of outer-product-based computing (\ref{op}) should guarantee that all {\em nonsquare} initial, intermediate, and final data tensors have to be always stationary within a 3D physical processor space to support chaining of processing,
whereas all {\em square} coefficient matrices have to be injected or streamed into processor space from the outside. This {\em admissible} mapping is achieved when
each stage of processing with a $n_s$-th specific 4D index space ($\mathcal{IS}_{n_s}$) is mapped to the {\em same} 3D processor space ($\mathcal{PS}$) along the corresponding directions of summation $n_s, s\in \{1,2,3\}$. 
These iteratively collapsed directions are assigned to the {\em discrete time}. 
Such three-stage 4D spatial ${\rightarrow}$ \{(3D spatial) + (1D time)\} mappings can be described for the selected tensor partition as follows:
\begin{subequations}
\renewcommand{\theequation}{\theparentequation.\arabic{equation}}
\label{lp}
\begin{equation}
{\text{Stage I.}}~~\mathcal{IS}_{n_3}: \langle n_1,n_2,n_3,k_3\rangle\in [N_1{\times}N_2{\times}N_3{\times N_3}]\xrightarrow[\mathbfcal X\rightarrow \mathbfcal{\dot X}]{\text{along} ~n_3}
\mathcal{PS}: \langle n_1,n_2,k_3\rangle\in [N_1{\times}N_2{\times}N_3]_{{\circlearrowright}N_3};
\label{lp1}
\end{equation}
\begin{equation}
\label{lp2}
{\text{Stage II.}}~~\mathcal{IS}_{n_1}: \langle n_1,n_2,k_3,k_1\rangle\in [N_1{\times}N_2{\times}N_3{\times N_1}]\xrightarrow[\mathbfcal {\dot X}\rightarrow \mathbfcal{\ddot X}]{\text{along} ~n_1}
\mathcal{PS}: \langle k_1,n_2,k_3\rangle\in[N_1{\times}N_2{\times}N_3]_{{\circlearrowright}N_1};
\end{equation}
\begin{equation}
\label{lp3}
{\text{Stage III.}}~\mathcal{IS}_{n_2}: \langle k_1,n_2,k_3,k_2\rangle\in [N_1{\times}N_2{\times}N_3{\times N_2}]\xrightarrow[\mathbfcal {\ddot X}\rightarrow \mathbfcal{\dddot X}]{\text{along} ~n_2}
\mathcal{PS}: \langle k_1,k_2,k_3\rangle\in[N_1{\times}N_2{\times}N_3]_{{\circlearrowright}N_2}.
\end{equation}
\end{subequations}

Here, each mapping collapses one dimension by zeroing the selected direction such that a 3D network of interconnected MAC cells implements the required $n_s$-summation by iteration along this direction $N_s$ times, shown above as ${\circlearrowright}N_s$, in the selected order of processing or summations $s=\{3,1,2\}$. 
All initial, intermediate and final $(N_1{\times}N_2{\times}N_3)$ data tensors remain inside the same $(N_1{\times}N_2{\times}N_3)$ processor space $\mathcal{PS}$, whereas all three slice-sharable coefficient matrices are injected or streamed into this $\mathcal{PS}$ from the three outside memories with $N_s$ channels each. 

Note that the same technique -- spatial mapping of an algorithm's iteration space to a processor space along the direction of summation -- has been previously used in the designing of a planar systolic array processor for the 2D discrete Fourier transform~\cite{289998} as well as in the
mapping of a 3D molecular dynamics problem into the Cerebras massively-parallel planar Wafer-Scale Engine~\cite{Santos2024BreakingTM}.  

\section{Acceleration of Trilinear Discrete Transforms}
\label{acceleration}

\subsection{A new kernel for outer-product-based GEMM implementation}
To our knowledge, there are two well-known output-stationary kernels of the outer-product-based GEMM implementation on a planar array processor:
\begin{enumerate}
\item[(1)] for the rectangular-by-rectangular matrix-matrix multiply-add  (RR-GEMM), where two input operand matrices are located in the separeted streaming memories outside of a {\em rectangular} array processor~\cite{agarwal_high-performance_1994};
\item[(2)] for the square-by-square matrix-matrix multiply-add (SS-GEMM),  where all three input and one output matrices are resided and reused inside of a {\em square} array processor~\cite{van_de_geijn_summa:_1997}.
\end{enumerate}
None of them is suitable for the acceleration of a {\em chaining} square-by-rectangular matrix-matrix multiply-add (SR-GEMM), where an output rectangular matrix is immediately used as an input matrix for the next stage of chain processing. Exactly this case is required for the acceleration of a 3D DXT and multilayer DNNs, where change-of-basis matrices and parameter/filter windows, respectively, are square while input and activation data are preferable rectangular. 

Here, the new also output-stationary SR-GEMM kernel is introduced:
\begin{enumerate}
\item[(3)] for the square-by-rectangular matrix-matrix multiply-add, where one input square matrix is located in a streaming memory outside of a planar array processor while other input and output rectangular matrices reside inside of this processor.
\end{enumerate}



The unified cellular $\mathcal{PS}$ obtained by linear mappings defines a Trilinear Algorithm/Accelerator Device Architecture (TriADA) as a fully distributed 3D network of atomic compute-storage-communication elements, or, simply, cells, interconnected by 3D crossover mesh of operand data lines (buses). This volumetric TriADA's part, which stores and processes the 3-mode data tensors, forms the Tensor Core.  The shareable mesh of horizontal $(\mathcal H)$, lateral $(\mathcal L)$, and frontal $(\mathcal F)$ data lines also connects the cells of the Tensor Core to three Decoupled Active Streaming Memories (DASM) or {\em Actuators} of three-stage processing. Each such directionally connected DASM stores and broadcasts on each time-step vector of the corresponding coefficient matrix to the specific planar face of the Tensor Core. Each operand line acts as a neuron's axon in the human brain, that is, transferring the activation signal or the same element of a vector from the neuron or the sender cell to all other connected to this axon or line neurons or cells. However, note that the axon of the biological neuron {\em uniquely} maintains the signal strength and receiving time to be practically the same for all, possible thousands, remotely connected neurons. For all this, the length of an axon can be many thousands of times longer than the size of the neuron.

The size of the volumetric Tensor Core $P_1{\times}P_2{\times}P_3$ is assumed to be larger than or equal to the size of the problem(s) with dimension $(N_1{\times}N_2{\times}N_3)$, that is, $P_s\geq N_s, s=1,2,3$. Otherwise, GEMM-like partitioning of the large problem into tiles or blocks should be considered. 
Each cell in the Tensor Core stores in local memory the corresponding element $x$ of an input tensor $\mathbfcal X_{N_1{\times}N_2{\times}N_3}$ as well as the elements $\dot x$, $\ddot x$, $\dddot x$ of the intermediate and final tensors $\mathbfcal{\dot X}_{N_1{\times}N_2{\times}N_3}$, $\mathbfcal{\ddot X}_{N_1{\times}N_2{\times}N_3}$ and $\mathbfcal{\dddot X}_{N_1{\times}N_2{\times}N_3}$, respectively, which are initially considered zero tensors. Following the above selected tensor partition, a horizontal slicing is initially used, where a 3D data grid is represented as $N_2$ data independent slices or $(N_1{\times}N_3)$-matrices, which are shown in Figure~\ref{S1} in light gray.

\subsection{Stage~I Acceleration}

In Stage~I, where summation is implemented along $n_3$-direction, see Eq.~(\ref{op1}) and mapping (\ref{lp1}), a rank-$N_3$ update is executed for each of $N_2$ horizontal slices by computing in each summation step a rank-1 update of an intermediate matrix/slice $\dot{\bf X}_{N_1{\times}N_3}^{(n_2)}$. Each rank-1 update is implemented as outer-product of an input vector-column ${\bf x}(n_3)_{N_1}^{(n_2)}\in {\bf X}_{N_1{\times}N_3}^{(n_2)}$ and a vector-row of coefficients ${\bf c}(n_3)_{N_3}\in {\bf C}_{N_3{\times}N_3},$ which is common for all $N_2$ horizontal slices or matrices as it is shown in  Eq.~(\ref{op1}). These $n_3$-rd vectors ${\bf x}(n_3)_{N_1}^{(n_2)}$ and ${\bf c}(n_3)_{N_3}$ are multicast along the
corresponding rows and columns of an intermediate matrix/slice $\dot{\bf X}_{N_1{\times}N_3}^{(n_2)}$, respectively. Note that because a coefficient matrix ${\bf C}_{N_3{\times}N_3}$ is square,  the size of a resulting matrix $\dot{\bf X}_{N_1{\times}N_3}^{(n_2)}$ is equal to the size of an input matrix ${\bf X}_{N_1{\times}N_3}^{(n_2)}$, i.e., both the input and the updated elements of these matrices can be stored and (re)used/updated locally in the corresponding cells of the Tensor Core.

\begin{figure}[htbp]
\begin{center}
\includegraphics[width=1\textwidth]{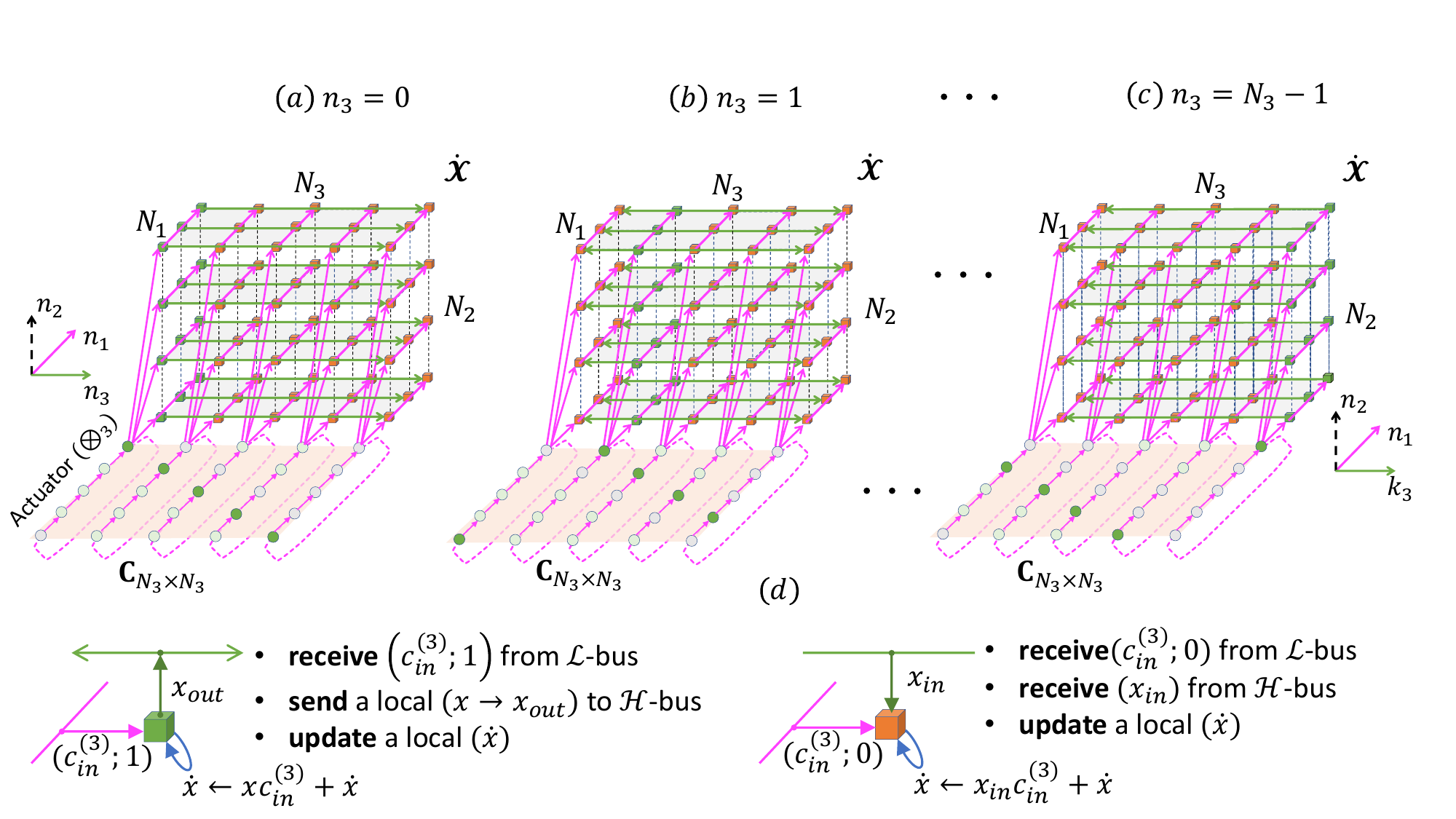}
\caption{Stage I: computing and data movement on each time-step $n_3\in[0,N_3)$, where $c_{in}^{(3)}$ is a corresponding element of a row-vector  ${\bf c}(n_3)_{N_3}\in {\bf C}_{N_3{\times}N_3}$, see Eq.~(\ref{op1}).
The rows of this single matrix is shared between all $N_2$ slices on each time-step.}
\label{S1}
\end{center}
\end{figure}

A coefficient matrix ${\bf C}_{N_3{\times}N_3}$ is stored in the so-called  Lateral Actuator~$({\otimes}_3)$ with $N_3$ streaming channels. This actuator starts the 3-mode tensor processing by injecting the $n_3$-rd vector-row ${\bf c}(n_3)_{N_3}$ into all Tensor Core cells using all lateral operand buses on the $(N_2{\times}N_3)$ front (back) face of the Tensor Core. This massively parallel single-step vector-to-matrix replication is shown in Figures~\ref{S1}$(a),(b),(c)$ in magenta. 
To activate the needed $n_3$-rd vector-column ${\bf x}(n_3)_{N_1}^{(n_2)}$ of a multiplicand matrix  ${\bf X}_{N_1{\times}N_3}^{(n_2)}$, all diagonal elements of a coefficient matrix ${\bf C}_{N_3{\times}N_3}$ are marked with a special tag=1 (shown for the Lateral Actuator in Figure~\ref{S1} as small green dots), whereas other elements are marked with a tag=0. Such column-row tag-based synchronization in matrix-by-matrix multiplication is only possible when one of the matrices is square, as in the case of orthogonal transformations. This arrangement makes activity of each cell independent on the cell's spatial location in TriADA network and, finally, on the problem size. However, in order to execute the generalized matrix-by-tensor multiplication (GEMT), which allows multiplication of non-square matrices, another (seems to be not trivial) coordinate-free synchronization policy should be used.

The coefficient vectors with tag=1 in position $n_3$, activate the needed in this time-step, the $n_3$-rd column of cells (shown as green cubes) in each horizontal slice. These pivotal cells activated by the input data are responsible for the local formation of an input vector-column ${\bf x}(n_3)_{N_1}^{(n_2)}\in {\bf X}_{N_1{\times}N_3}^{(n_2)}$ that is sent along horizontal operand data buses to other cells connected to these buses (shown as orange cubes). The crossover transferred vector-row and vector-column are used (in this casting order) to implement a single outer-product, i.e., correctly update an intermediate matrix within a slice. These successive data-driven rank-1 updates are shown in Figure~\ref{S1} ({\it a}), ({\it b}) and ({\it c}) for the first time-step $n_3=0$, the second time-step $n_3=1$ and the last time-step $n_3=N_3-1$, respectively.
Each streamed tagged vector-row of a coefficient matrix activates in each time-step a relevant vector-column of ``green'' cells, making the whole processing independent of the problem size. 
In general, due to the fact that a sum of rank-1 updates can be executed in arbitrary order, any not-overlapped position of the tag=1 in a streamed coefficient vector is admissible (not only for diagonal elements).

Figure~\ref{S1}$(d)$ shows cell activity in each time-step for the ``green" cells
(activated by the input data element $c_{in}$ with tag=1) and the ``orange" cells (activated by the data element $c_{in}$ with tag=0).  A data-driven and index-free formulation of the cell's local processing (data communication and computing) on each time-step is used. This formulation allows cell activity to be independent of the size of the problem. That is, each cell in the TriADA network does not "feel" how big or small the problem is, making the cell's activity totally under control of the input data. This independency of cell activity makes the TriADA network highly scalable and adaptable to the size and shape of the input data tensor.

In total, $N_3$ time-steps are needed to compute Stage~I, assuming that all sends, receives, and updates are implemented in the Tensor Core with $N_1{\times}N_2{\times}N_3$ cells, simultaneously, in one time-step. In terms of data streams, this massively parallel implementation of Stage~I corresponds to the $\dot{\mathbfcal X}$-stationary variant when the result of each MAC operation is stored and updated locally in each cell. The processing of this stage is totally under the control of the streaming vectors of a diagonally tagged coefficient matrix ${\bf C}_{N_3{\times}N_3}$ stored in the Lateral Actuator~$(\otimes_3)$. After completion of stage, that is, after multicasting of  the last vector-row of coefficients, Actuator~$({\otimes}_3)$ transfers control to the Horizontal Actuator~$({\otimes}_1)$, which starts the Stage~II.

\subsection{Stage~II Acceleration}
\begin{figure}[htbp]
\begin{center}
\includegraphics[width=1\textwidth]{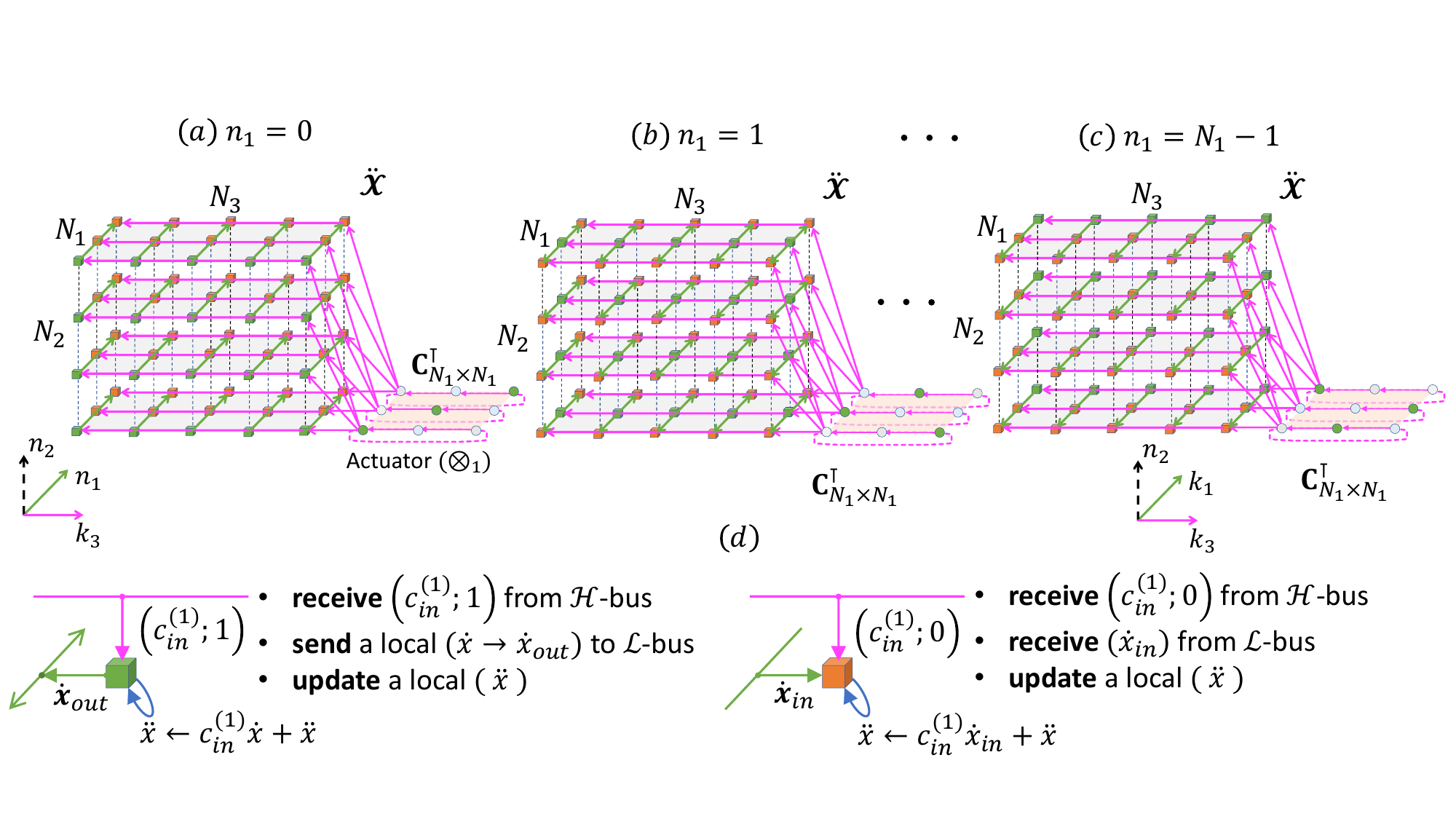}
\caption{Stage II: computing and data movement on each of $N_1$ time-steps, where $c_{in}^{(1)}$ is a corresponding element of a vector-column  ${\bf c}(n_1)_{N_1}\in {\bf C}_{N_1{\times}N_1}^\top$, see Eq.~(\ref{op2}).}
\label{S2}
\end{center}
\end{figure}

On Stage~II, the summation along the tensor direction ${n}_1\in [0,N_1)$ is implemented following Eq.~(\ref{op2}) and mapping~(\ref{lp2}). 
Each cell in the Tensor Core has a previously calculated element $\dot{x}\in \dot {\bf X}_{N_1{\times}N_3}^{(n_2)}$ and an element $\ddot{x}\in \ddot {\bf X}_{N_1{\times}N_3}^{(n_2)},$ that is initially equal to zero. Like before, in each step of summation, the same vector-column 
${\bf c}(n_1)_{N_1}\in {\bf C}_{N_1{\times}N_1}^\top$ is replicated to all $n_2=[0,N_2)$ slices of cells that store matrices $\ddot {\bf X}_{N_1{\times}N_3}^{(n_2)}$. This replication of a vector on the planar $(N_1{\times}N_2)$ right (left) face of the Tensor Core is shown in Figure~\ref{S2} in magenta. Like in Stage~I, the diagonal elements of a coefficient matrix ${\bf C}_{N_1{\times}N_1}^\top$, which is stored in the so-called Horizontal Actuator $(\otimes_1)$, are tagged. The other vector-row $\dot{\bf x}(n_1)_{N_3}^{(n_2)}\in \dot {\bf X}_{N_1{\times}N_3}^{(n_2)}$ is formed inside of a $n_2$-nd slice by $n_1$-st column of "green" cells, which multicast this vector via horizonal operand buses. These horizontal buses are shown in Figure~\ref{S2} in green. 

Figures~\ref{S2}$(a),(b),(c)$ show the activity of cells in the Tensor Core for the first $n_1=0$, the second $n_1=1$, and the last $n_1=N_1-1$ time-steps during the calculation of the intermediate tensor $\ddot{\mathbfcal X}$. Like in the previous stage, a stream memory for storing and cyclically transferring tagged vectors of the coefficient matrix ${\bf C}_{N_1{\times}N_1}^\top$ also works as a multihead drum memory. 
Figure~\ref{S2}$(d)$ illustrates a data-driven activity of "green" and "orange" cells in each time-step. Each cell starts MAC computing upon readiness of the three operand data. 

Note that a linear projection of the 3D $\mathcal PS$ or Tensor Core along the direction $n_2$ for both Stage~I and II, gives a planar $(N_1{\times}N_3)$ array processor with vector PEs or vector cells. Each vector PE executes in each time-step the {\texttt{SAXPY}} or vector update operation with an input scalar coefficient and $N_2$-length vectors with a proper vector data exchange. Therefore, a planar array of vector PEs is able to extremely accelerate the execution of a bilinear transform of tensor data. However, the trilinear transform requires summation and, therefore, access to the tensor data along the missing in the bilinear transform direction.  

In total, TriADA spends $N_1$ time-steps and produces $N_1{\times}N_2{\times}N_3$ MAC results per time-step to compute Stage~II.
It is also computed as the multiple of output- or $\ddot{\mathbfcal X}$-stationary matrix-by-matrix multiplications, that is, an intermediate tensor $\ddot{\mathbfcal X}$ remains inside the Tensor Core cellular network. After finishing this stage, the Actuator $(\otimes_1)$ transfers control to the next Actuator $(\otimes_2)$ which starts the last Stage~III. 

\subsection{Stage~III Acceleration}

Stage~III, which is based on the summation by the access to tensor data along the dimension $n_2\in [0,N_2),$ see Eq.~(\ref{op3}) and mapping (\ref{lp3}), can be implemented by repartitioning an intermediate tensor $\ddot{\mathbfcal X}_{N_1{\times}N_2{\times}N_3}$ into frontal or lateral slices that include a desirable index $n_2$, see Figures~\ref{partition}(b),(c). 
Here, a lateral slicing of the tensor is used following the equality (\ref{repartition}).
After finishing Stage~II, each cell has a computed element $\ddot{x}\in \ddot{\mathbfcal X}_{N_1{\times}N_2{\times}N_3}$ and an element $\dddot{ x}\in \dddot{\mathbfcal X}_{N_1{\times}N_2{\times}N_3}$ that is initially equal to zero.

\begin{figure}[htbp]
\begin{center}
\includegraphics[width=1\textwidth]{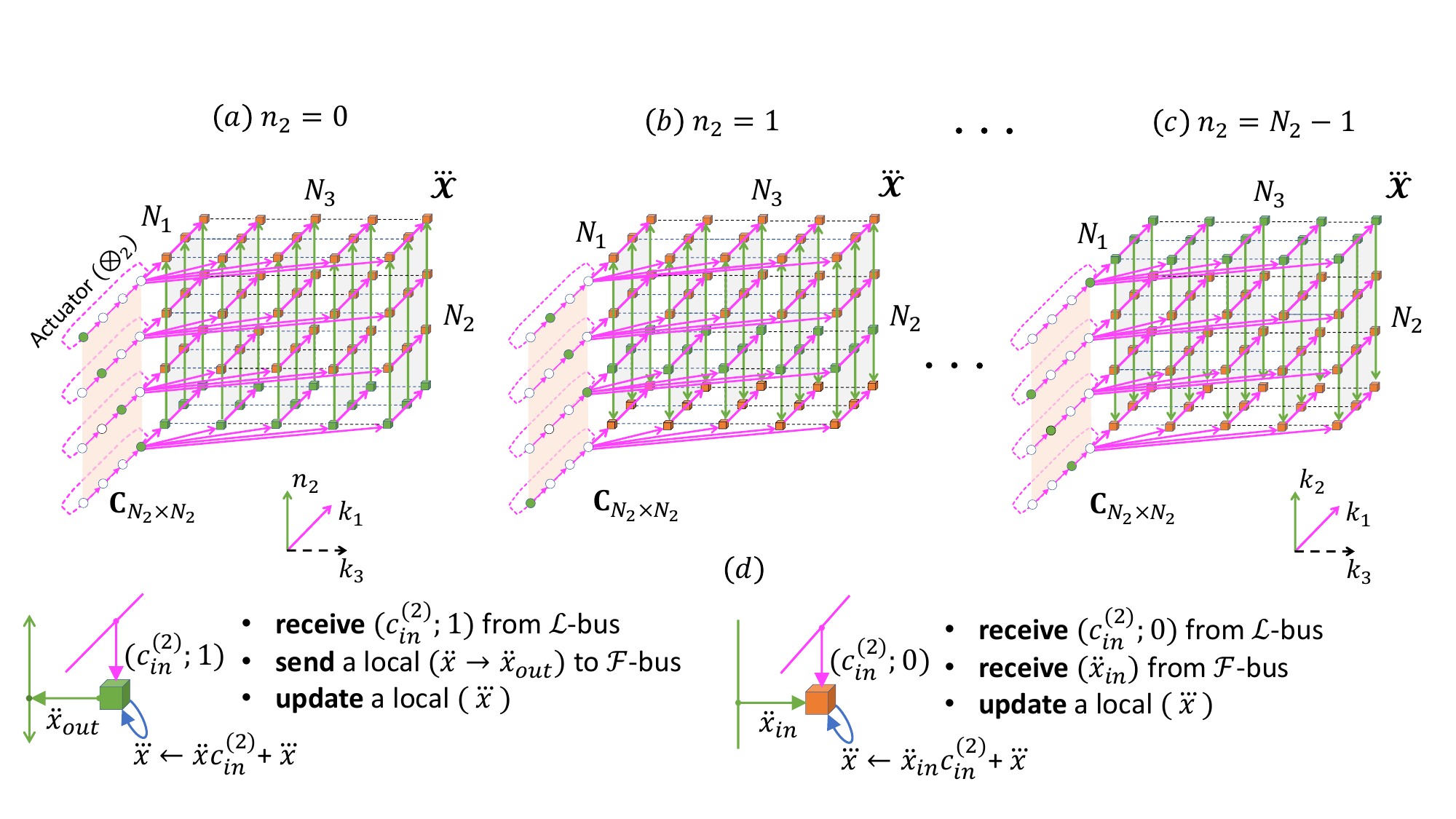}
\caption{Stage III: computing and data movement on each of $N_2$ steps,  where $c_{in}^{(2)}$ is a corresponding element of a vector  ${\bf c}(n_2)_{N_2}\in {\bf C}_{N_2{\times}N_2}$, see Eq.~(\ref{op3}).}
\label{S3}
\end{center}
\end{figure}

After receiving a control, Actuator~$(\otimes_2)$ starts the final stage of processing by streaming the first row-vector ${\bf c}(n_2=0)_{N_2}\in {\bf C}_{N_2{\times}N_2}$ along the lateral operand buses to all $k_3 \in [0,N_3)$ slices of cells, which store the corresponding lateral matrices $\ddot {\bf X}_{N_1{\times}N_2}^{(k_3)}$.  This vector-to-matrix multicast, oriented to the frontal $(N_2{\times}N_3)$ face of the tensor, is shown in Figures~\ref{S3} in magenta. 
In each time-step, a rank-1 update is computed by multicasting of a row-vector ${\bf c}(n_2)_{N_2}\in {\bf C}_{N_2{\times}N_2}$ and a column-vector ${\bf{\ddot x}}(n_2)^{(k_3)}_{N_1}\in \ddot {\bf X}_{N_1{\times}N_2}^{(k_3)}$ to update a final matrix 
$\dddot {\bf X}_{N_1{\times}N_2}^{(k_3)}$. 
As above, multicast of a column-vector ${\bf{\ddot x}}(n_2)^{(k_3)}_{N_1}$ is activated by corresponding "green" cells under the control of an outside streamed and tagged row-vector ${\bf c}(n_2)_{N_2}$.

Figures~\ref{S3}$(a),(b),(c)$ show the lateral tensor partition and the data movement in a TriADA for the first $n_2=0$, the second $n_2=1$, and the last $n_2=N_2-1$ time-steps during the computation of a final tensor $\dddot{\mathbfcal X}_{N_1{\times}N_2{\times}N_3}$. Figure~\ref{S3}$(d)$ shows a data-driven activity of "green" and "orange" cells in each time-step. As above, a cell activity does not depend on the problem size. In total, $N_2$ time-steps are needed to compute a Stage~III. This stage is also computed as the multiple of data-independent output- or $\dddot{\mathbfcal X}$-stationary matrix-matrix multiplications which remain inside of the TriADA network. 

As a result, all $(N_1 N_2  N_3)(N_1{+} N_2{+} N_3)$ MAC operations,  required to compute the discrete trilinear $(N_1{\times}N_2{\times}N_3)$ forward or inverse transformation, can be executed on the TriADA with $N_1{\times}N_2{\times}N_3$ crossover interconnected cells in the linear number of $N_1{+}N_2{+}N_3$ time-steps with 100\% efficiency. However, this level of efficiency is only possible for the dense GEMT variant. Sparse matrix and/or tensor data can drastically affect the efficiency of processing. Of course, it is possible to avoid the execution of MAC or update operation by checking operands on zero inside each cell in TriADA, but it assumes sending and receiving zero-valued elements together with valid data. In the following, a more flexible and less energy-consuming method is proposed.

\section{Management of Unstructured Data Sparsity}
In today's AI systems, data sparsity, i.e. the presence of many zero or insignificant values, is a key feature that helps make models more efficient and manageable. AI workloads, especially deep learning, often exhibit high levels of sparsity, typically ranging from 50\% to more than 90\%. This means that in many models, 50\% to more than 90\% of the data or parameters might be zero or insignificant, which helps reduce computational costs and memory use.
Sparsity is crucial because it allows AI models to run faster and use less energy. The management of computing and the movement of sparse data continues to evolve, balancing efficiency gains with accuracy, making sparsity a critical aspect of modern AI development~\cite{hoefler2021sparsitydeeplearningpruning, yen2022s4highsparsityhighperformanceai, Wu_2023, Xue_2023, thangarasa2024sparseiftsparseisofloptransformations}. At the same time, the more general unstructured sparsity is considered to be not adequate for acceleration DNN models on existing parallel hardware, such as GPUs~\cite{jeong2025enablingunstructuredsparseacceleration}.

In the following, a new method is proposed to manage processing of unstructured sparse tensor data. This method avoids not only redundant computing, but also communication of zero-valued elements in the TriADA network. We call this approach of saving arithmetic and communication operations in processing unstructured sparse data as the Elastic Sparse Outer-product Processing (ESOP) method. Unstructured sparsity refers to the case in which zero-valued elements are randomly scattered across structured data sets such as vectors, matrices, and tensors. The ESOP method is based on the distinctive feature of outer-product operation to form partial updates of an intermediate matrix by crossing coordinate elements of two operand vectors such that intersection of zero-valued vector elements does not change all elements in the corresponding row and/or column of a matrix. The ESOP method equally supports both static and dynamic data sparsity. 

\begin{figure}[htbp]
    \begin{center}
    \includegraphics[width=9cm]{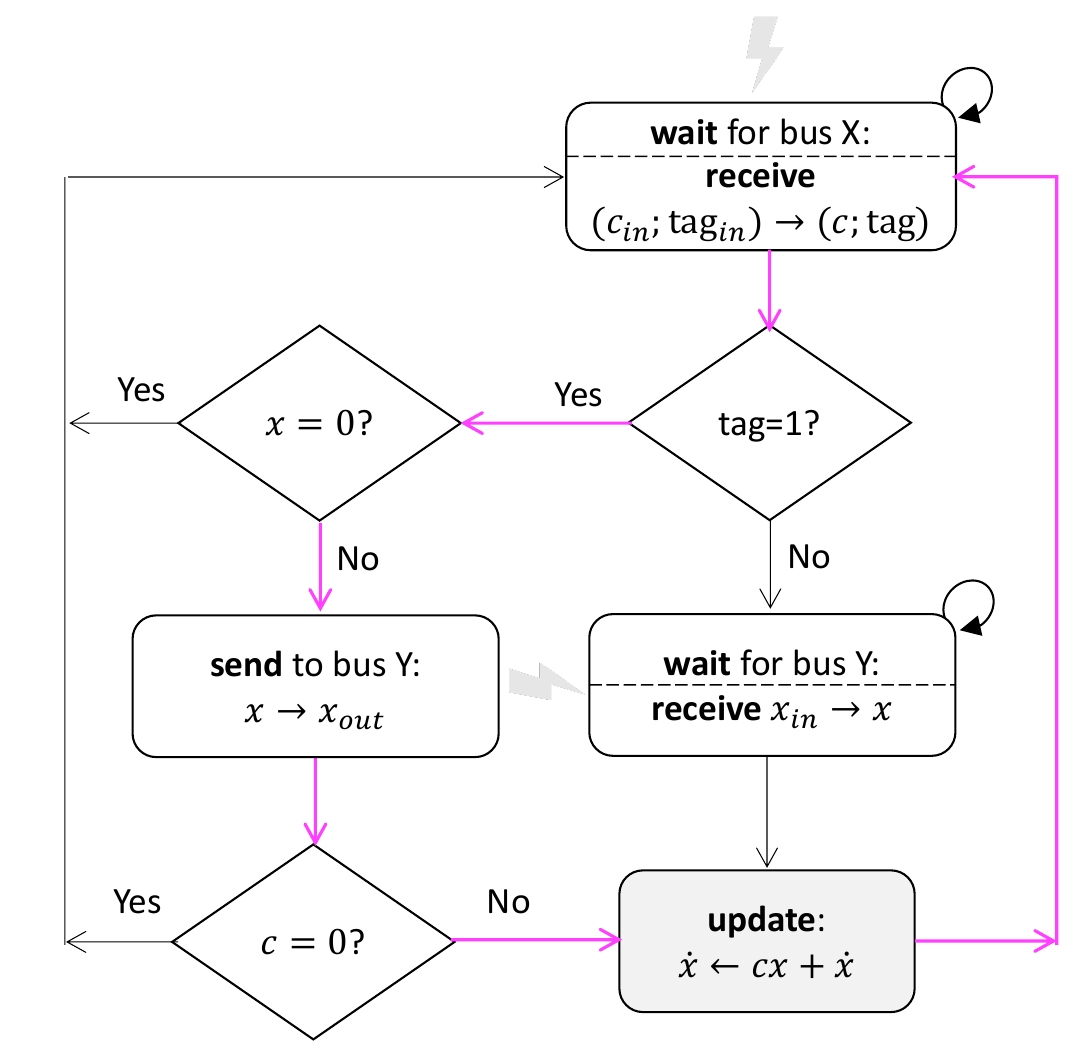}
    \caption{Diagram of cell's activity to manage a data sparsity.}
    \label{sparsity}
    \end{center}
\end{figure}

The cell activity diagram in each time-step is shown in Figure~\ref{sparsity}, where the critical or longest processing path is colored magenta.
In Figure~\ref{sparsity}, the pair of orthogonal {\tt{(X,Y)}}-buses is either {\tt{(L,H)}} for Stage~I or {\tt{(H,L)}} for Stage~II or {\tt{(L,F)}} for Stage~III,
where {\tt{L}}, {\tt{H}} and {\tt{F}} designate the Lateral, Horizontal, and Frontal operand buses, respectively.
In the proposed scenario, the pipeline period of the actuator data stream should not be less than the execution time of the critical path by the cell.
Therefore, if the data stream has not zero-valued vectors, the total number of needed for computing linear transformation time-steps is always equal to the length of current summation, although the number of multiply-and-add or update operations in sparse data computing can be sufficiently less than that of the dense case. The critical path of the local processing defines the time-step period, which is common or global for all cells in the TriADA network. The proposed ESOP method dynamically supports in each time-step all possible combinations of vector data sparsity: dense-dense, sparse-dense, dense-sparse, and sparse-sparse, totally avoiding computing with zero-valued vectors. 

The zero-valued non-diagonal or non-pivotal coefficients $(c=0;{\text{tag}}=0)$ are never sent by the actuator to {\tt{X}} buses. 
Moreover, a zero-vector has ${\text{tag}}_{in}=0$ for all elements, including pivot, such that the actuator skips sending this all-zero vector to {\tt{X}} buses, saving one time-step and related dynamic energy consumption of the array processing.  
In each time-step, the cell receives only a single element $(c_{in}\neq0;{\text{tag}}_{in}=1)$ or $(c_{in}=0;{\text{tag}}_{in}=1)$ or $(c_{in}\neq0;{\text{tag}}_{in}=0)$ from the corresponding {{\tt X}}-bus, which has been sent to this operand bus by the actuator. All cells connected to {\tt{X}}-bus and receiving the ${\text{tag}}_{in}=1$ are responsible for forming and sending the elements of a vector $\bf x$ through a set of orthogonal {\tt{Y}}-buses. However, cells with zero-valued local elements $x=0$ did not send them, forcing all other cells connected to these {\tt{Y}}-buses to remain in the waiting state, thus avoiding unnecessary updates of local data. This waiting will be canceled by receiving a new coefficient $(c_{in}\neq 0;{\text{tag}_{in}}=0)$ from the {\tt{X}}-bus. Depending on the data size and degree of data sparsity, this strategy may sufficiently reduce the dynamic energy consumption related to massively sending/receiving and updating tensor data.

Moreover, the proposed ESOP method permits not only reducing the energy consumption of processing but also increasing the accuracy and stability of computing. In fact, computing the linear transformation includes an accumulation operation, where the previous result is used as input for subsequent steps. Because the rounding or roundoff error is also propagated and accumulated together with the partial sum, the final rounding error depends on the total number of local data update steps or the length of the calculation. The ESOP approach avoids the update or multiply-add operations with zero-valued operands within each cell and, therefore, reduces the length of the calculation. The more sparse the data, the more arithmetic and communication operations are avoided, improving, as a result, the accuracy of the computing, and collectively decreasing the total dynamic energy consumption of parallel processing. Obviously, computing different data elements may consume different amounts of energy. 

\section{Conclusion}

The proposed TriADA approach integrates novel algorithmic advancements with a specialized distributed hardware architecture, with the aim of fundamentally mitigate computational, memory, and energy constraints. TriADA's transformative capabilities are rooted in four core innovations:
\begin{itemize}
    \item A massively parallel, low-rank, cellular algorithm for the fast and efficient computation of a family of trilinear (or 3-dimensional) discrete ortogonal transformations (3D-DXTs). This algorithm is a specific case of the more general 3-mode matrix-by-tensor (3D-GEMT) multiplication.
    \item A new kernel for outer-product-based GEMM computing, featuring decoupled streaming memory, specifically designed to accelerate 3D-GEMT operation.
    \item An isomorphic to the proposed algorithm with a new GEMM kernel fully distributed 3D network of atomic computing-storage-communication elements, or simply cells. The cells are interconnected by a 3D crossover mesh of operand data lines (buses), which are also linked to three decoupled streaming memories.
    \item The "Elastic Sparse Outer Product" method, an intellegent approach that avoids unnecessary computing and communication operations on zero-valued operands, thereby enhancing energy efficiency, computational accuracy, and stability for sparse data processing. 
\end{itemize}
More specifically, the proposed data-driven Trilinear Algorithm and isomorphic Device Architecture (TriADA) executes three-dimensional $(N_1{\times}N_2{\times}N_3)$ transformations in a linear $N_1{+}N_2{+}N_3$ number of time steps.
The same $\langle P_1{\times}P_2{\times}P_3\rangle$ TriADA network can be used to store and accelerate the solution of any $(N_1{\times}N_2{\times}N_3)$ problem for which $N_s\leq P_s$, $s=1,2,3$. 
The low-rank GEMT algorithm and TriADA architecture are well suited to accelerate multilinear tensor operations, which are the most compute-intensive parts of advanced AI and HPC workloads. 

The proposed massively parallel low-rank algorithm and data-driven TriADA architecture are simple, regular, homogeneous, scalable, and highly data-reusable, making them ready for the upcoming trillion-transistor era. We envision this paper as a foundational step in exploring the potential of an extremely distributed, highly scalable, tensor computing and storage network with many millions of mesh-interconnected simple compute-storage-communication cells. 

\section{Acknowledgments}
This work was partially supported by the Competitive Research Funds of the University of Aizu. 

\bibliographystyle{ACM-Reference-Format}
\bibliography{cite_list}


\begin{thebibliography}{70}


\ifx \showCODEN    \undefined \def \showCODEN     #1{\unskip}     \fi
\ifx \showDOI      \undefined \def \showDOI       #1{#1}\fi
\ifx \showISBNx    \undefined \def \showISBNx     #1{\unskip}     \fi
\ifx \showISBNxiii \undefined \def \showISBNxiii  #1{\unskip}     \fi
\ifx \showISSN     \undefined \def \showISSN      #1{\unskip}     \fi
\ifx \showLCCN     \undefined \def \showLCCN      #1{\unskip}     \fi
\ifx \shownote     \undefined \def \shownote      #1{#1}          \fi
\ifx \showarticletitle \undefined \def \showarticletitle #1{#1}   \fi
\ifx \showURL      \undefined \def \showURL       {\relax}        \fi
\providecommand\bibfield[2]{#2}
\providecommand\bibinfo[2]{#2}
\providecommand\natexlab[1]{#1}
\providecommand\showeprint[2][]{arXiv:#2}

\bibitem[Acar et~al\mbox{.}(2009)]%
        {Acar2009}
\bibfield{author}{\bibinfo{person}{Evrim Acar}, \bibinfo{person}{Robert Harrison}, \bibinfo{person}{Frank Olken}, \bibinfo{person}{Orly Alter}, \bibinfo{person}{Manal Helal}, \bibinfo{person}{Larsson Omberg}, \bibinfo{person}{Brett Bader}, \bibinfo{person}{Anthony Kennedy}, \bibinfo{person}{Haesun Park}, \bibinfo{person}{Zhaojun Bai}, \bibinfo{person}{Dongmin Kim}, \bibinfo{person}{Robert Plemmons}, \bibinfo{person}{Gregory Beylkin}, \bibinfo{person}{Tamara Kolda}, \bibinfo{person}{Stefan Ragnarsson}, \bibinfo{person}{Lieven Delathauwer}, \bibinfo{person}{Julien Langou}, \bibinfo{person}{Sri Priya}, \bibinfo{person}{Ponnapalli Dhillon}, {and} \bibinfo{person}{Charles Loan}.} \bibinfo{year}{2009}\natexlab{}.
\newblock \bibinfo{booktitle}{\emph{Workshop Report Future Directions in Tensor-Based Computation and Modeling}}.
\newblock \bibinfo{type}{{T}echnical {R}eport}. \bibinfo{institution}{NDF}. \bibinfo{pages}{20} pages.
\newblock
\urldef\tempurl%
\url{https://doi.org/10.13140/2.1.4040.4807}
\showDOI{\tempurl}


\bibitem[Agarwal et~al\mbox{.}(1994)]%
        {agarwal_high-performance_1994}
\bibfield{author}{\bibinfo{person}{R.~C. Agarwal}, \bibinfo{person}{F.~G. Gustavson}, {and} \bibinfo{person}{M. Zubair}.} \bibinfo{year}{1994}\natexlab{}.
\newblock \showarticletitle{A high-performance matrix-multiplication algorithm on a distributed-memory parallel computer, using overlapped communication}.
\newblock \bibinfo{journal}{\emph{IBM Journal of Research and Development}} \bibinfo{volume}{38}, \bibinfo{number}{6} (\bibinfo{date}{Nov.} \bibinfo{year}{1994}), \bibinfo{pages}{673--681}.
\newblock
\showISSN{0018-8646, 0018-8646}
\urldef\tempurl%
\url{https://doi.org/10.1147/rd.386.0673}
\showDOI{\tempurl}


\bibitem[Alam et~al\mbox{.}(2024)]%
        {electronics13152988}
\bibfield{author}{\bibinfo{person}{Shahanur Alam}, \bibinfo{person}{Chris Yakopcic}, \bibinfo{person}{Qing Wu}, \bibinfo{person}{Mark Barnell}, \bibinfo{person}{Simon Khan}, {and} \bibinfo{person}{Tarek~M. Taha}.} \bibinfo{year}{2024}\natexlab{}.
\newblock \showarticletitle{Survey of Deep Learning Accelerators for Edge and Emerging Computing}.
\newblock \bibinfo{journal}{\emph{Electronics}} \bibinfo{volume}{13}, \bibinfo{number}{15} (\bibinfo{year}{2024}), \bibinfo{pages}{1--44}.
\newblock
\showISSN{2079-9292}
\urldef\tempurl%
\url{https://doi.org/10.3390/electronics13152988}
\showDOI{\tempurl}


\bibitem[Baumgartner et~al\mbox{.}(2005)]%
        {1386652}
\bibfield{author}{\bibinfo{person}{G. Baumgartner}, \bibinfo{person}{A. Auer}, \bibinfo{person}{D.E. Bernholdt}, \bibinfo{person}{A. Bibireata}, \bibinfo{person}{V. Choppella}, \bibinfo{person}{D. Cociorva}, \bibinfo{person}{Xiaoyang Gao}, \bibinfo{person}{R.J. Harrison}, \bibinfo{person}{S. Hirata}, \bibinfo{person}{S. Krishnamoorthy}, \bibinfo{person}{S. Krishnan}, \bibinfo{person}{Chi chung Lam}, \bibinfo{person}{Qingda Lu}, \bibinfo{person}{M. Nooijen}, \bibinfo{person}{R.M. Pitzer}, \bibinfo{person}{J. Ramanujam}, \bibinfo{person}{P. Sadayappan}, {and} \bibinfo{person}{A. Sibiryakov}.} \bibinfo{year}{2005}\natexlab{}.
\newblock \showarticletitle{Synthesis of High-Performance Parallel Programs for a Class of ab Initio Quantum Chemistry Models}.
\newblock \bibinfo{journal}{\emph{Proc. IEEE}} \bibinfo{volume}{93}, \bibinfo{number}{2} (\bibinfo{year}{2005}), \bibinfo{pages}{276--292}.
\newblock
\urldef\tempurl%
\url{https://doi.org/10.1109/JPROC.2004.840311}
\showDOI{\tempurl}


\bibitem[Bavikadi et~al\mbox{.}(2022)]%
        {9739030}
\bibfield{author}{\bibinfo{person}{Sathwika Bavikadi}, \bibinfo{person}{Abhijitt Dhavlle}, \bibinfo{person}{Amlan Ganguly}, \bibinfo{person}{Anand Haridass}, \bibinfo{person}{Hagar Hendy}, \bibinfo{person}{Cory Merkel}, \bibinfo{person}{Vijay~Janapa Reddi}, \bibinfo{person}{Purab~Ranjan Sutradhar}, \bibinfo{person}{Arun Joseph}, {and} \bibinfo{person}{Sai~Manoj Pudukotai~Dinakarrao}.} \bibinfo{year}{2022}\natexlab{}.
\newblock \showarticletitle{A Survey on Machine Learning Accelerators and Evolutionary Hardware Platforms}.
\newblock \bibinfo{journal}{\emph{IEEE Design \& Test}} \bibinfo{volume}{39}, \bibinfo{number}{3} (\bibinfo{year}{2022}), \bibinfo{pages}{91--116}.
\newblock
\urldef\tempurl%
\url{https://doi.org/10.1109/MDAT.2022.3161126}
\showDOI{\tempurl}


\bibitem[Bowers et~al\mbox{.}(2006)]%
        {bowers_scalable_2006}
\bibfield{author}{\bibinfo{person}{Kevin~J. Bowers}, \bibinfo{person}{David~E. Chow}, \bibinfo{person}{Huafeng Xu}, \bibinfo{person}{Ron~O. Dror}, \bibinfo{person}{Michael~P. Eastwood}, \bibinfo{person}{Brent~A. Gregersen}, \bibinfo{person}{John~L. Klepeis}, \bibinfo{person}{Istvan Kolossvary}, \bibinfo{person}{Mark~A. Moraes}, \bibinfo{person}{Federico~D. Sacerdoti}, \bibinfo{person}{John~K. Salmon}, \bibinfo{person}{Yibing Shan}, {and} \bibinfo{person}{David~E. Shaw}.} \bibinfo{year}{2006}\natexlab{}.
\newblock \showarticletitle{Scalable {Algorithms} for {Molecular} {Dynamics} {Simulations} on {Commodity} {Clusters}}. In \bibinfo{booktitle}{\emph{{ACM}/{IEEE} {SC} 2006 {Conference} ({SC}'06)}}. \bibinfo{publisher}{IEEE}, \bibinfo{address}{Tampa, FL}, \bibinfo{pages}{43--43}.
\newblock
\urldef\tempurl%
\url{https://doi.org/10.1109/SC.2006.54}
\showDOI{\tempurl}


\bibitem[Cannon(1969)]%
        {cannon_cellular_1969}
\bibfield{author}{\bibinfo{person}{Lynn~Elliot Cannon}.} \bibinfo{year}{1969}\natexlab{}.
\newblock \emph{\bibinfo{title}{A cellular computer to implement the {Kalman} filter algorithm}}.
\newblock Ph.{D}. dissertation. \bibinfo{school}{Montana State University}.
\newblock
\urldef\tempurl%
\url{https://www.proquest.com/openview/a64b8bf2c8ba4cc59bae06592eb9cf17/1.pdf?pq-origsite=gscholar&cbl=18750&diss=y}
\showURL{%
\tempurl}


\bibitem[Cooley and Tukey(1965)]%
        {CooleyTukey}
\bibfield{author}{\bibinfo{person}{James Cooley} {and} \bibinfo{person}{John Tukey}.} \bibinfo{year}{1965}\natexlab{}.
\newblock \showarticletitle{An Algorithm for the Machine Calculation of Complex Fourier Series}.
\newblock \bibinfo{journal}{\emph{Math. Comp.}} \bibinfo{volume}{19}, \bibinfo{number}{90} (\bibinfo{year}{1965}), \bibinfo{pages}{297--301}.
\newblock


\bibitem[de~Carvalho et~al\mbox{.}(2022)]%
        {9796646}
\bibfield{author}{\bibinfo{person}{João P.~L. de Carvalho}, \bibinfo{person}{José~E. Moreira}, {and} \bibinfo{person}{José~Nelson Amaral}.} \bibinfo{year}{2022}\natexlab{}.
\newblock \showarticletitle{Compiling for the IBM Matrix Engine for Enterprise Workloads}.
\newblock \bibinfo{journal}{\emph{IEEE Micro}} \bibinfo{volume}{42}, \bibinfo{number}{5} (\bibinfo{year}{2022}), \bibinfo{pages}{34--40}.
\newblock
\urldef\tempurl%
\url{https://doi.org/10.1109/MM.2022.3176529}
\showDOI{\tempurl}


\bibitem[De~Lathauwer et~al\mbox{.}(2000)]%
        {de_lathauwer_multilinear_2000}
\bibfield{author}{\bibinfo{person}{Lieven De~Lathauwer}, \bibinfo{person}{Bart De~Moor}, {and} \bibinfo{person}{Joos Vandewalle}.} \bibinfo{year}{2000}\natexlab{}.
\newblock \showarticletitle{A {Multilinear} {Singular} {Value} {Decomposition}}.
\newblock \bibinfo{journal}{\emph{SIAM J. Matrix Anal. Appl.}} \bibinfo{volume}{21}, \bibinfo{number}{4} (\bibinfo{date}{Jan.} \bibinfo{year}{2000}), \bibinfo{pages}{1253--1278}.
\newblock
\showISSN{0895-4798, 1095-7162}
\urldef\tempurl%
\url{https://doi.org/10.1137/S0895479896305696}
\showDOI{\tempurl}


\bibitem[Di et~al\mbox{.}(2025)]%
        {di2025surveyerrorboundedlossycompression}
\bibfield{author}{\bibinfo{person}{Sheng Di}, \bibinfo{person}{Jinyang Liu}, \bibinfo{person}{Kai Zhao}, \bibinfo{person}{Xin Liang}, \bibinfo{person}{Robert Underwood}, \bibinfo{person}{Zhaorui Zhang}, \bibinfo{person}{Milan Shah}, \bibinfo{person}{Yafan Huang}, \bibinfo{person}{Jiajun Huang}, \bibinfo{person}{Xiaodong Yu}, \bibinfo{person}{Congrong Ren}, \bibinfo{person}{Hanqi Guo}, \bibinfo{person}{Grant Wilkins}, \bibinfo{person}{Dingwen Tao}, \bibinfo{person}{Jiannan Tian}, \bibinfo{person}{Sian Jin}, \bibinfo{person}{Zizhe Jian}, \bibinfo{person}{Daoce Wang}, \bibinfo{person}{MD~Hasanur Rahman}, \bibinfo{person}{Boyuan Zhang}, \bibinfo{person}{Shihui Song}, \bibinfo{person}{Jon~C. Calhoun}, \bibinfo{person}{Guanpeng Li}, \bibinfo{person}{Kazutomo Yoshii}, \bibinfo{person}{Khalid~Ayed Alharthi}, {and} \bibinfo{person}{Franck Cappello}.} \bibinfo{year}{2025}\natexlab{}.
\newblock \bibinfo{title}{A Survey on Error-Bounded Lossy Compression for Scientific Datasets}.
\newblock
\newblock
\showeprint[arxiv]{2404.02840}~[cs.DC]
\urldef\tempurl%
\url{https://arxiv.org/abs/2404.02840}
\showURL{%
\tempurl}


\bibitem[Domke et~al\mbox{.}(2021)]%
        {domke2021matrixengineshighperformance}
\bibfield{author}{\bibinfo{person}{Jens Domke}, \bibinfo{person}{Emil Vatai}, \bibinfo{person}{Aleksandr Drozd}, \bibinfo{person}{Peng Chen}, \bibinfo{person}{Yosuke Oyama}, \bibinfo{person}{Lingqi Zhang}, \bibinfo{person}{Shweta Salaria}, \bibinfo{person}{Daichi Mukunoki}, \bibinfo{person}{Artur Podobas}, \bibinfo{person}{Mohamed Wahib}, {and} \bibinfo{person}{Satoshi Matsuoka}.} \bibinfo{year}{2021}\natexlab{}.
\newblock \bibinfo{title}{Matrix Engines for High Performance Computing:A Paragon of Performance or Grasping at Straws?}
\newblock
\newblock
\showeprint[arxiv]{2010.14373}~[cs.DC]
\urldef\tempurl%
\url{https://arxiv.org/abs/2010.14373}
\showURL{%
\tempurl}


\bibitem[Dongarra and Sullivan(2000)]%
        {814652}
\bibfield{author}{\bibinfo{person}{J. Dongarra} {and} \bibinfo{person}{F. Sullivan}.} \bibinfo{year}{2000}\natexlab{}.
\newblock \showarticletitle{Guest Editors Introduction to the top 10 algorithms}.
\newblock \bibinfo{journal}{\emph{Computing in Science \& Engineering}} \bibinfo{volume}{2}, \bibinfo{number}{1} (\bibinfo{year}{2000}), \bibinfo{pages}{22--23}.
\newblock
\urldef\tempurl%
\url{https://doi.org/10.1109/MCISE.2000.814652}
\showDOI{\tempurl}


\bibitem[Dubey et~al\mbox{.}(2022)]%
        {10.1016/j.neucom.2022.06.111}
\bibfield{author}{\bibinfo{person}{Shiv~Ram Dubey}, \bibinfo{person}{Satish~Kumar Singh}, {and} \bibinfo{person}{Bidyut~Baran Chaudhuri}.} \bibinfo{year}{2022}\natexlab{}.
\newblock \showarticletitle{Activation functions in deep learning: A comprehensive survey and benchmark}.
\newblock \bibinfo{journal}{\emph{Neurocomput.}} \bibinfo{volume}{503}, \bibinfo{number}{C} (\bibinfo{date}{Sept.} \bibinfo{year}{2022}), \bibinfo{pages}{92–108}.
\newblock
\showISSN{0925-2312}
\urldef\tempurl%
\url{https://doi.org/10.1016/j.neucom.2022.06.111}
\showDOI{\tempurl}


\bibitem[Dutta et~al\mbox{.}(2019)]%
        {8758338}
\bibfield{author}{\bibinfo{person}{Sanghamitra Dutta}, \bibinfo{person}{Viveck Cadambe}, {and} \bibinfo{person}{Pulkit Grover}.} \bibinfo{year}{2019}\natexlab{}.
\newblock \showarticletitle{{“Short-Dot”}: Computing Large Linear Transforms Distributedly Using Coded Short Dot Products}.
\newblock \bibinfo{journal}{\emph{IEEE Transactions on Information Theory}} \bibinfo{volume}{65}, \bibinfo{number}{10} (\bibinfo{year}{2019}), \bibinfo{pages}{6171--6193}.
\newblock
\urldef\tempurl%
\url{https://doi.org/10.1109/TIT.2019.2927558}
\showDOI{\tempurl}


\bibitem[Fang et~al\mbox{.}(2021)]%
        {fang_accelerating_2021}
\bibfield{author}{\bibinfo{person}{Chao Fang}, \bibinfo{person}{Liulu He}, \bibinfo{person}{Haonan Wang}, \bibinfo{person}{Jinghe Wei}, {and} \bibinfo{person}{Zhongfeng Wang}.} \bibinfo{year}{2021}\natexlab{}.
\newblock \showarticletitle{Accelerating {3D} {Convolutional} {Neural} {Networks} {Using} {3D} {Fast} {Fourier} {Transform}}. In \bibinfo{booktitle}{\emph{2021 {IEEE} {International} {Symposium} on {Circuits} and {Systems} ({ISCAS})}}. \bibinfo{publisher}{IEEE}, \bibinfo{address}{Daegu, Korea}, \bibinfo{pages}{1--5}.
\newblock
\showISBNx{9781728192017}
\urldef\tempurl%
\url{https://doi.org/10.1109/ISCAS51556.2021.9401765}
\showDOI{\tempurl}


\bibitem[Fox et~al\mbox{.}(1987)]%
        {fox_matrix_1987}
\bibfield{author}{\bibinfo{person}{G.C Fox}, \bibinfo{person}{S.W Otto}, {and} \bibinfo{person}{A.J.G Hey}.} \bibinfo{year}{1987}\natexlab{}.
\newblock \showarticletitle{Matrix algorithms on a hypercube {I}: {Matrix} multiplication}.
\newblock \bibinfo{journal}{\emph{Parallel Comput.}} \bibinfo{volume}{4}, \bibinfo{number}{1} (\bibinfo{date}{Feb.} \bibinfo{year}{1987}), \bibinfo{pages}{17--31}.
\newblock
\showISSN{01678191}
\urldef\tempurl%
\url{https://doi.org/10.1016/0167-8191(87)90060-3}
\showDOI{\tempurl}


\bibitem[Gemini(2025)]%
        {gemini_gemm_notation_report}
\bibfield{author}{\bibinfo{person}{Google Gemini}.} \bibinfo{year}{2025}\natexlab{}.
\newblock \bibinfo{title}{Comparative Analysis of Notations for Accelerating General Matrix-Matrix Multiplication}.
\newblock \bibinfo{howpublished}{Generated by Google Gemini, stored on Google Drive}.
\newblock
\urldef\tempurl%
\url{https://docs.google.com/document/d/11ZGtIVbeBVFuKhDv-jFAKq-Jnaa6vvPerFATJIt-ALc/edit?usp=sharing}
\showURL{%
\tempurl}
\newblock
\shownote{Report generated in response to prompt: "Please provide a comparative analysis of the dot-product, outer-product, and SAXPY notations for accelerating general matrix-matrix multiplication (GEMM) on modern hardware."}.


\bibitem[Golub and Van~Loan(2013)]%
        {golub_matrix_2013}
\bibfield{author}{\bibinfo{person}{Gene~H. Golub} {and} \bibinfo{person}{Charles~F. Van~Loan}.} \bibinfo{year}{2013}\natexlab{}.
\newblock \bibinfo{booktitle}{\emph{Matrix {Computations}} (\bibinfo{edition}{4th} ed.)}.
\newblock \bibinfo{publisher}{The Johns Hopkins University Press}, \bibinfo{address}{Baltimore, MD}.
\newblock
\showISBNx{1421407949, 9781421407944}
\urldef\tempurl%
\url{https://doi.org/10.56021/9781421407944}
\showURL{%
\tempurl}


\bibitem[Goto and Van De~Geijn(2008)]%
        {goto_anatomy_2008}
\bibfield{author}{\bibinfo{person}{Kazushige Goto} {and} \bibinfo{person}{Robert~A. Van De~Geijn}.} \bibinfo{year}{2008}\natexlab{}.
\newblock \showarticletitle{Anatomy of high-performance matrix multiplication}.
\newblock \bibinfo{journal}{\emph{ACM Trans. Math. Software}} \bibinfo{volume}{34}, \bibinfo{number}{3} (\bibinfo{date}{May} \bibinfo{year}{2008}), \bibinfo{pages}{1--25}.
\newblock
\showISSN{0098-3500, 1557-7295}
\urldef\tempurl%
\url{https://doi.org/10.1145/1356052.1356053}
\showDOI{\tempurl}


\bibitem[Gustavson(1978)]%
        {10.1145/355791.355796}
\bibfield{author}{\bibinfo{person}{Fred~G. Gustavson}.} \bibinfo{year}{1978}\natexlab{}.
\newblock \showarticletitle{Two Fast Algorithms for Sparse Matrices: Multiplication and Permuted Transposition}.
\newblock \bibinfo{journal}{\emph{ACM Trans. Math. Softw.}} \bibinfo{volume}{4}, \bibinfo{number}{3} (\bibinfo{date}{Sept.} \bibinfo{year}{1978}), \bibinfo{pages}{250--269}.
\newblock
\showISSN{0098-3500}
\urldef\tempurl%
\url{https://doi.org/10.1145/355791.355796}
\showDOI{\tempurl}


\bibitem[{H. Kung}(1982)]%
        {kung_why_1982}
\bibfield{author}{\bibinfo{person}{{H. Kung}}.} \bibinfo{year}{1982}\natexlab{}.
\newblock \showarticletitle{Why systolic architectures?}
\newblock \bibinfo{journal}{\emph{Computer}} \bibinfo{volume}{15}, \bibinfo{number}{1} (\bibinfo{date}{Jan.} \bibinfo{year}{1982}), \bibinfo{pages}{37--46}.
\newblock
\showISSN{0018-9162}
\urldef\tempurl%
\url{https://doi.org/10.1109/MC.1982.1653825}
\showDOI{\tempurl}


\bibitem[He et~al\mbox{.}(2025)]%
        {he2025waferllmwaferscalellminference}
\bibfield{author}{\bibinfo{person}{Congjie He}, \bibinfo{person}{Yeqi Huang}, \bibinfo{person}{Pei Mu}, \bibinfo{person}{Ziming Miao}, \bibinfo{person}{Jilong Xue}, \bibinfo{person}{Lingxiao Ma}, \bibinfo{person}{Fan Yang}, {and} \bibinfo{person}{Luo Mai}.} \bibinfo{year}{2025}\natexlab{}.
\newblock \bibinfo{title}{WaferLLM: A Wafer-Scale LLM Inference System}.
\newblock
\newblock
\showeprint[arxiv]{2502.04563}~[cs.LG]
\urldef\tempurl%
\url{https://arxiv.org/abs/2502.04563}
\showURL{%
\tempurl}


\bibitem[Hendrycks and Gimpel(2023)]%
        {hendrycks2023gaussianerrorlinearunits}
\bibfield{author}{\bibinfo{person}{Dan Hendrycks} {and} \bibinfo{person}{Kevin Gimpel}.} \bibinfo{year}{2023}\natexlab{}.
\newblock \bibinfo{title}{Gaussian Error Linear Units (GELUs)}.
\newblock
\newblock
\showeprint[arxiv]{1606.08415}~[cs.LG]
\urldef\tempurl%
\url{https://arxiv.org/abs/1606.08415}
\showURL{%
\tempurl}


\bibitem[Hoefler et~al\mbox{.}(2021)]%
        {hoefler2021sparsitydeeplearningpruning}
\bibfield{author}{\bibinfo{person}{Torsten Hoefler}, \bibinfo{person}{Dan Alistarh}, \bibinfo{person}{Tal Ben-Nun}, \bibinfo{person}{Nikoli Dryden}, {and} \bibinfo{person}{Alexandra Peste}.} \bibinfo{year}{2021}\natexlab{}.
\newblock \bibinfo{title}{Sparsity in Deep Learning: Pruning and growth for efficient inference and training in neural networks}.
\newblock
\newblock
\showeprint[arxiv]{2102.00554}~[cs.LG]
\urldef\tempurl%
\url{https://arxiv.org/abs/2102.00554}
\showURL{%
\tempurl}


\bibitem[Hu et~al\mbox{.}(2024)]%
        {10460211}
\bibfield{author}{\bibinfo{person}{Yang Hu}, \bibinfo{person}{Xinhan Lin}, \bibinfo{person}{Huizheng Wang}, \bibinfo{person}{Zhen He}, \bibinfo{person}{Xingmao Yu}, \bibinfo{person}{Jiahao Zhang}, \bibinfo{person}{Qize Yang}, \bibinfo{person}{Zheng Xu}, \bibinfo{person}{Sihan Guan}, \bibinfo{person}{Jiahao Fang}, \bibinfo{person}{Haoran Shang}, \bibinfo{person}{Xinru Tang}, \bibinfo{person}{Xu Dai}, \bibinfo{person}{Shaojun Wei}, {and} \bibinfo{person}{Shouyi Yin}.} \bibinfo{year}{2024}\natexlab{}.
\newblock \showarticletitle{Wafer-Scale Computing: Advancements, Challenges, and Future Perspectives [Feature]}.
\newblock \bibinfo{journal}{\emph{IEEE Circuits and Systems Magazine}} \bibinfo{volume}{24}, \bibinfo{number}{1} (\bibinfo{year}{2024}), \bibinfo{pages}{52--81}.
\newblock
\urldef\tempurl%
\url{https://doi.org/10.1109/MCAS.2024.3349669}
\showDOI{\tempurl}


\bibitem[Hübner et~al\mbox{.}(2025)]%
        {hübner2025applevsorangesevaluating}
\bibfield{author}{\bibinfo{person}{Paul Hübner}, \bibinfo{person}{Andong Hu}, \bibinfo{person}{Ivy Peng}, {and} \bibinfo{person}{Stefano Markidis}.} \bibinfo{year}{2025}\natexlab{}.
\newblock \bibinfo{title}{Apple vs. Oranges: Evaluating the Apple Silicon M-Series SoCs for HPC Performance and Efficiency}.
\newblock
\newblock
\showeprint[arxiv]{2502.05317}~[cs.AR]
\urldef\tempurl%
\url{https://arxiv.org/abs/2502.05317}
\showURL{%
\tempurl}


\bibitem[Ikegaki et~al\mbox{.}(2011)]%
        {ikegaki_3d-dct_2011}
\bibfield{author}{\bibinfo{person}{Yuki Ikegaki}, \bibinfo{person}{Toshiaki Miyazaki}, {and} \bibinfo{person}{Stanislav~G. Sedukhin}.} \bibinfo{year}{2011}\natexlab{}.
\newblock \showarticletitle{{3D}-{DCT} {Processor} and {Its} {FPGA} {Implementation}}.
\newblock \bibinfo{journal}{\emph{IEICE Transactions on Information and Systems}} \bibinfo{volume}{E94-D}, \bibinfo{number}{7} (\bibinfo{year}{2011}), \bibinfo{pages}{1409--1418}.
\newblock
\showISSN{0916-8532, 1745-1361}
\urldef\tempurl%
\url{https://doi.org/10.1587/transinf.E94.D.1409}
\showDOI{\tempurl}


\bibitem[Jeong et~al\mbox{.}(2025)]%
        {jeong2025enablingunstructuredsparseacceleration}
\bibfield{author}{\bibinfo{person}{Geonhwa Jeong}, \bibinfo{person}{Po-An Tsai}, \bibinfo{person}{Abhimanyu~R. Bambhaniya}, \bibinfo{person}{Stephen~W. Keckler}, {and} \bibinfo{person}{Tushar Krishna}.} \bibinfo{year}{2025}\natexlab{}.
\newblock \bibinfo{title}{Enabling Unstructured Sparse Acceleration on Structured Sparse Accelerators}.
\newblock
\newblock
\showeprint[arxiv]{2403.07953}~[cs.LG]
\urldef\tempurl%
\url{https://arxiv.org/abs/2403.07953}
\showURL{%
\tempurl}


\bibitem[Jouppi et~al\mbox{.}(2023)]%
        {jouppi2023tpuv4opticallyreconfigurable}
\bibfield{author}{\bibinfo{person}{Norman~P. Jouppi}, \bibinfo{person}{George Kurian}, \bibinfo{person}{Sheng Li}, \bibinfo{person}{Peter Ma}, \bibinfo{person}{Rahul Nagarajan}, \bibinfo{person}{Lifeng Nai}, \bibinfo{person}{Nishant Patil}, \bibinfo{person}{Suvinay Subramanian}, \bibinfo{person}{Andy Swing}, \bibinfo{person}{Brian Towles}, \bibinfo{person}{Cliff Young}, \bibinfo{person}{Xiang Zhou}, \bibinfo{person}{Zongwei Zhou}, {and} \bibinfo{person}{David Patterson}.} \bibinfo{year}{2023}\natexlab{}.
\newblock \bibinfo{title}{TPU v4: An Optically Reconfigurable Supercomputer for Machine Learning with Hardware Support for Embeddings}.
\newblock
\newblock
\showeprint[arxiv]{2304.01433}~[cs.AR]
\urldef\tempurl%
\url{https://arxiv.org/abs/2304.01433}
\showURL{%
\tempurl}


\bibitem[Kolda and Bader(2009)]%
        {kolda_tensor_2009}
\bibfield{author}{\bibinfo{person}{Tamara~G. Kolda} {and} \bibinfo{person}{Brett~W. Bader}.} \bibinfo{year}{2009}\natexlab{}.
\newblock \showarticletitle{Tensor {Decompositions} and {Applications}}.
\newblock \bibinfo{journal}{\emph{SIAM Rev.}} \bibinfo{volume}{51}, \bibinfo{number}{3} (\bibinfo{date}{Aug.} \bibinfo{year}{2009}), \bibinfo{pages}{455--500}.
\newblock
\showISSN{0036-1445, 1095-7200}
\urldef\tempurl%
\url{https://doi.org/10.1137/07070111X}
\showDOI{\tempurl}


\bibitem[Kung and Leiserson(1985)]%
        {kung__systolic_1985}
\bibfield{author}{\bibinfo{person}{H. Kung} {and} \bibinfo{person}{C. Leiserson}.} \bibinfo{year}{1985}\natexlab{}.
\newblock \bibinfo{title}{Systolic array apparatuses for matrix computations}.
\newblock
\newblock
\urldef\tempurl%
\url{https://patents.google.com/patent/US4493048A/}
\showURL{%
\tempurl}
\newblock
\shownote{Patent US 4,493,048}.


\bibitem[Li et~al\mbox{.}(1997)]%
        {li_poly_1997}
\bibfield{author}{\bibinfo{person}{J. Li}, \bibinfo{person}{A. Skjellum}, {and} \bibinfo{person}{R.~D. Falgout}.} \bibinfo{year}{1997}\natexlab{}.
\newblock \showarticletitle{A poly-algorithm for parallel dense matrix multiplication on two-dimensional process grid topologies}.
\newblock \bibinfo{journal}{\emph{Concurrency: Practice and Experience}} \bibinfo{volume}{9}, \bibinfo{number}{5} (\bibinfo{year}{1997}), \bibinfo{pages}{345--389}.
\newblock
\urldef\tempurl%
\url{https://doi.org/10.1002/(SICI)1096-9128(199705)9:5<345::AID-CPE258>3.0.CO;2-7}
\showDOI{\tempurl}


\bibitem[Li et~al\mbox{.}(2023a)]%
        {pdpu23}
\bibfield{author}{\bibinfo{person}{Qiong Li}, \bibinfo{person}{Chao Fang}, {and} \bibinfo{person}{Zhongfeng Wang}.} \bibinfo{year}{2023}\natexlab{a}.
\newblock \showarticletitle{{PDPU}: An Open-Source Posit Dot-Product Unit for Deep Learning Applications}. In \bibinfo{booktitle}{\emph{2023 IEEE International Symposium on Circuits and Systems (ISCAS)}}. \bibinfo{publisher}{IEEE}, \bibinfo{address}{Monterey, CA, USA}, \bibinfo{pages}{1--5}.
\newblock
\urldef\tempurl%
\url{https://doi.org/10.1109/ISCAS46773.2023.10182007}
\showDOI{\tempurl}


\bibitem[Li et~al\mbox{.}(2023b)]%
        {10139817}
\bibfield{author}{\bibinfo{person}{Shiqing Li}, \bibinfo{person}{Shuo Huai}, {and} \bibinfo{person}{Weichen Liu}.} \bibinfo{year}{2023}\natexlab{b}.
\newblock \showarticletitle{An Efficient Gustavson-Based Sparse Matrix–Matrix Multiplication Accelerator on Embedded FPGAs}.
\newblock \bibinfo{journal}{\emph{IEEE Transactions on Computer-Aided Design of Integrated Circuits and Systems}} \bibinfo{volume}{42}, \bibinfo{number}{12} (\bibinfo{year}{2023}), \bibinfo{pages}{4671--4680}.
\newblock
\urldef\tempurl%
\url{https://doi.org/10.1109/TCAD.2023.3281719}
\showDOI{\tempurl}


\bibitem[Lim(2021)]%
        {lim_tensors_2021}
\bibfield{author}{\bibinfo{person}{Lek-Heng Lim}.} \bibinfo{year}{2021}\natexlab{}.
\newblock \showarticletitle{Tensors in computations}.
\newblock \bibinfo{journal}{\emph{Acta Numerica}}  \bibinfo{volume}{30} (\bibinfo{date}{May} \bibinfo{year}{2021}), \bibinfo{pages}{555--764}.
\newblock
\showISSN{0962-4929, 1474-0508}
\urldef\tempurl%
\url{https://doi.org/10.1017/S0962492921000076}
\showDOI{\tempurl}


\bibitem[Lin et~al\mbox{.}(2018)]%
        {lin_fft-based_2018}
\bibfield{author}{\bibinfo{person}{Sheng Lin}, \bibinfo{person}{Ning Liu}, \bibinfo{person}{Mahdi Nazemi}, \bibinfo{person}{Hongjia Li}, \bibinfo{person}{Caiwen Ding}, \bibinfo{person}{Yanzhi Wang}, {and} \bibinfo{person}{Massoud Pedram}.} \bibinfo{year}{2018}\natexlab{}.
\newblock \showarticletitle{{FFT}-based deep learning deployment in embedded systems}. In \bibinfo{booktitle}{\emph{2018 {Design}, {Automation} \& {Test} in {Europe} {Conference} \& {Exhibition} ({DATE})}}. \bibinfo{publisher}{IEEE}, \bibinfo{address}{Dresden, Germany}, \bibinfo{pages}{1045--1050}.
\newblock
\showISBNx{9783981926309}
\urldef\tempurl%
\url{https://doi.org/10.23919/DATE.2018.8342166}
\showDOI{\tempurl}


\bibitem[Lirkov(2020)]%
        {lirkov_performance_2020}
\bibfield{author}{\bibinfo{person}{Ivan Lirkov}.} \bibinfo{year}{2020}\natexlab{}.
\newblock \showarticletitle{Performance {Analysis} of a {Scalable} {Algorithm} for {3D} {Linear} {Transforms} on {Supercomputer} with {Intel} {Processors}/{Co}-{Processors}}.
\newblock \bibinfo{journal}{\emph{Cybernetics and Information Technologies}} \bibinfo{volume}{20}, \bibinfo{number}{6} (\bibinfo{date}{Dec.} \bibinfo{year}{2020}), \bibinfo{pages}{94--104}.
\newblock
\showISSN{1314-4081}
\urldef\tempurl%
\url{https://doi.org/10.2478/cait-2020-0064}
\showDOI{\tempurl}


\bibitem[Liu and Wong(2024)]%
        {10589682}
\bibfield{author}{\bibinfo{person}{Mark Liu} {and} \bibinfo{person}{H.-S.~Philip Wong}.} \bibinfo{year}{2024}\natexlab{}.
\newblock \showarticletitle{The Path to a 1-Trillion-Transistor GPU: AI's Boom Demands New Chip Technology}.
\newblock \bibinfo{journal}{\emph{IEEE Spectrum}} \bibinfo{volume}{61}, \bibinfo{number}{7} (\bibinfo{year}{2024}), \bibinfo{pages}{22--27}.
\newblock
\urldef\tempurl%
\url{https://doi.org/10.1109/MSPEC.2024.10589682}
\showDOI{\tempurl}


\bibitem[Liu and Parhi(2023)]%
        {10190238}
\bibfield{author}{\bibinfo{person}{Xingyi Liu} {and} \bibinfo{person}{Keshab~K. Parhi}.} \bibinfo{year}{2023}\natexlab{}.
\newblock \showarticletitle{Tensor Decomposition for Model Reduction in Neural Networks: A Review}.
\newblock \bibinfo{journal}{\emph{IEEE Circuits and Systems Magazine}} \bibinfo{volume}{23}, \bibinfo{number}{2} (\bibinfo{year}{2023}), \bibinfo{pages}{8--28}.
\newblock
\urldef\tempurl%
\url{https://doi.org/10.1109/MCAS.2023.3267921}
\showURL{%
\tempurl}


\bibitem[Lu et~al\mbox{.}(2012)]%
        {LU2012338}
\bibfield{author}{\bibinfo{person}{Qingda Lu}, \bibinfo{person}{Xiaoyang Gao}, \bibinfo{person}{Sriram Krishnamoorthy}, \bibinfo{person}{Gerald Baumgartner}, \bibinfo{person}{J. Ramanujam}, {and} \bibinfo{person}{P. Sadayappan}.} \bibinfo{year}{2012}\natexlab{}.
\newblock \showarticletitle{Empirical performance model-driven data layout optimization and library call selection for tensor contraction expressions}.
\newblock \bibinfo{journal}{\emph{J. Parallel and Distrib. Comput.}} \bibinfo{volume}{72}, \bibinfo{number}{3} (\bibinfo{year}{2012}), \bibinfo{pages}{338--352}.
\newblock
\showISSN{0743-7315}
\urldef\tempurl%
\url{https://doi.org/10.1016/j.jpdc.2011.09.006}
\showDOI{\tempurl}


\bibitem[Luo et~al\mbox{.}(2024)]%
        {luo2024benchmarkingdissectingnvidiahopper}
\bibfield{author}{\bibinfo{person}{Weile Luo}, \bibinfo{person}{Ruibo Fan}, \bibinfo{person}{Zeyu Li}, \bibinfo{person}{Dayou Du}, \bibinfo{person}{Qiang Wang}, {and} \bibinfo{person}{Xiaowen Chu}.} \bibinfo{year}{2024}\natexlab{}.
\newblock \bibinfo{title}{Benchmarking and Dissecting the Nvidia Hopper GPU Architecture}.
\newblock
\newblock
\showeprint[arxiv]{2402.13499}~[cs.AR]
\urldef\tempurl%
\url{https://arxiv.org/abs/2402.13499}
\showURL{%
\tempurl}


\bibitem[Malapally et~al\mbox{.}(2023)]%
        {malapally_scalability_2023}
\bibfield{author}{\bibinfo{person}{Nitin Malapally}, \bibinfo{person}{Viacheslav Bolnykh}, \bibinfo{person}{Estela Suarez}, \bibinfo{person}{Paolo Carloni}, \bibinfo{person}{Thomas Lippert}, {and} \bibinfo{person}{Davide Mandelli}.} \bibinfo{year}{2023}\natexlab{}.
\newblock \bibinfo{title}{Scalability of {3D}-{DFT} by block tensor-matrix multiplication on the {JUWELS} {Cluster}}.
\newblock
\newblock
\urldef\tempurl%
\url{http://arxiv.org/abs/2303.13337}
\showURL{%
\tempurl}
\newblock
\shownote{arXiv:2303.13337 [physics]}.


\bibitem[Malapally et~al\mbox{.}(2024)]%
        {MALAPALLY2024104945}
\bibfield{author}{\bibinfo{person}{Nitin Malapally}, \bibinfo{person}{Viacheslav Bolnykh}, \bibinfo{person}{Estela Suarez}, \bibinfo{person}{Paolo Carloni}, \bibinfo{person}{Thomas Lippert}, {and} \bibinfo{person}{Davide Mandelli}.} \bibinfo{year}{2024}\natexlab{}.
\newblock \showarticletitle{{3D DFT by block tensor-matrix multiplication via a modified Cannon's algorithm: Implementation and scaling on distributed-memory clusters with fat tree networks}}.
\newblock \bibinfo{journal}{\emph{J. Parallel and Distrib. Comput.}}  \bibinfo{volume}{193} (\bibinfo{year}{2024}), \bibinfo{pages}{104945}.
\newblock
\showISSN{0743-7315}
\urldef\tempurl%
\url{https://doi.org/10.1016/j.jpdc.2024.104945}
\showDOI{\tempurl}


\bibitem[Orenes-Vera et~al\mbox{.}(2023)]%
        {10.1145/3577193.3593708}
\bibfield{author}{\bibinfo{person}{Marcelo Orenes-Vera}, \bibinfo{person}{Ilya Sharapov}, \bibinfo{person}{Robert Schreiber}, \bibinfo{person}{Mathias Jacquelin}, \bibinfo{person}{Philippe Vandermersch}, {and} \bibinfo{person}{Sharan Chetlur}.} \bibinfo{year}{2023}\natexlab{}.
\newblock \showarticletitle{Wafer-Scale Fast Fourier Transforms}. In \bibinfo{booktitle}{\emph{Proceedings of the 37th ACM International Conference on Supercomputing}} (Orlando, FL, USA) \emph{(\bibinfo{series}{ICS '23})}. \bibinfo{publisher}{Association for Computing Machinery}, \bibinfo{address}{New York, NY, USA}, \bibinfo{pages}{180–191}.
\newblock
\showISBNx{9798400700569}
\urldef\tempurl%
\url{https://doi.org/10.1145/3577193.3593708}
\showDOI{\tempurl}


\bibitem[Pilz et~al\mbox{.}(2025)]%
        {pilz2025trendsaisupercomputers}
\bibfield{author}{\bibinfo{person}{Konstantin~F. Pilz}, \bibinfo{person}{James Sanders}, \bibinfo{person}{Robi Rahman}, {and} \bibinfo{person}{Lennart Heim}.} \bibinfo{year}{2025}\natexlab{}.
\newblock \bibinfo{title}{Trends in AI Supercomputers}.
\newblock
\newblock
\showeprint[arxiv]{2504.16026}~[cs.CY]
\urldef\tempurl%
\url{https://arxiv.org/abs/2504.16026}
\showURL{%
\tempurl}


\bibitem[Pogue and Nicolici(2024)]%
        {10323219}
\bibfield{author}{\bibinfo{person}{T.~E. Pogue} {and} \bibinfo{person}{N. Nicolici}.} \bibinfo{year}{2024}\natexlab{}.
\newblock \showarticletitle{Fast Inner-Product Algorithms and Architectures for Deep Neural Network Accelerators}.
\newblock \bibinfo{journal}{\emph{IEEE Trans. Comput.}} \bibinfo{volume}{73}, \bibinfo{number}{02} (\bibinfo{date}{feb} \bibinfo{year}{2024}), \bibinfo{pages}{495--509}.
\newblock
\showISSN{1557-9956}
\urldef\tempurl%
\url{https://doi.org/10.1109/TC.2023.3334140}
\showDOI{\tempurl}


\bibitem[Rutledge and {Jouan-Rimbaud Bouveresse}(2007)]%
        {RUTLEDGE2007170}
\bibfield{author}{\bibinfo{person}{Douglas~N. Rutledge} {and} \bibinfo{person}{Delphine {Jouan-Rimbaud Bouveresse}}.} \bibinfo{year}{2007}\natexlab{}.
\newblock \showarticletitle{Multi-way analysis of outer product arrays using {PARAFAC}}.
\newblock \bibinfo{journal}{\emph{Chemometrics and Intelligent Laboratory Systems}} \bibinfo{volume}{85}, \bibinfo{number}{2} (\bibinfo{year}{2007}), \bibinfo{pages}{170--178}.
\newblock
\showISSN{0169-7439}
\urldef\tempurl%
\url{https://doi.org/10.1016/j.chemolab.2006.06.011}
\showDOI{\tempurl}


\bibitem[Sakai and Sedukhin(2013)]%
        {sakai_3d_nodate}
\bibfield{author}{\bibinfo{person}{Tomoya Sakai} {and} \bibinfo{person}{Stanislav Sedukhin}.} \bibinfo{year}{2013}\natexlab{}.
\newblock \bibinfo{booktitle}{\emph{{3D} {Discrete} {Transforms} with {Cubical} {Data} {Decomposition} on the {IBM} {Blue} {Gene}/{Q}}}.
\newblock University of Aizu, Japan.
\newblock
\urldef\tempurl%
\url{https://u-aizu.ac.jp/files/page/research/techreport/2013-001.pdf}
\showURL{%
\tempurl}


\bibitem[Santos et~al\mbox{.}(2024)]%
        {Santos2024BreakingTM}
\bibfield{author}{\bibinfo{person}{Kylee Santos}, \bibinfo{person}{Stan Moore}, \bibinfo{person}{Tomas Oppelstrup}, \bibinfo{person}{Amirali Sharifian}, \bibinfo{person}{Ilya Sharapov}, \bibinfo{person}{Aidan Thompson}, \bibinfo{person}{Delyan~Z. Kalchev}, \bibinfo{person}{Danny Perez}, \bibinfo{person}{Robert Schreiber}, \bibinfo{person}{Scott Pakin}, \bibinfo{person}{Edgar~A. Leon}, \bibinfo{person}{James~H. Laros}, \bibinfo{person}{Michael James}, {and} \bibinfo{person}{Sivasankaran Rajamanickam}.} \bibinfo{year}{2024}\natexlab{}.
\newblock \showarticletitle{Breaking the Molecular Dynamics Timescale Barrier Using a Wafer-Scale System}. In \bibinfo{booktitle}{\emph{SC24: International Conference for High Performance Computing, Networking, Storage and Analysis}}. \bibinfo{publisher}{IEEE}, \bibinfo{address}{Atlanta, GA, USA}, \bibinfo{pages}{1--13}.
\newblock
\urldef\tempurl%
\url{https://doi.org/10.1109/SC41406.2024.00014}
\showDOI{\tempurl}


\bibitem[Schieffer et~al\mbox{.}(2024)]%
        {10590025}
\bibfield{author}{\bibinfo{person}{Gabin Schieffer}, \bibinfo{person}{Daniel~Araújo De~Medeiros}, \bibinfo{person}{Jennifer Faj}, \bibinfo{person}{Aniruddha Marathe}, {and} \bibinfo{person}{Ivy Peng}.} \bibinfo{year}{2024}\natexlab{}.
\newblock \showarticletitle{On the Rise of AMD Matrix Cores: Performance, Power Efficiency, and Programmability}. In \bibinfo{booktitle}{\emph{2024 IEEE International Symposium on Performance Analysis of Systems and Software (ISPASS)}}. \bibinfo{publisher}{IEEE}, \bibinfo{address}{Indianapolis, IN, USA}, \bibinfo{pages}{132--143}.
\newblock
\urldef\tempurl%
\url{https://doi.org/10.1109/ISPASS61541.2024.00022}
\showDOI{\tempurl}


\bibitem[Sedukhin(1994)]%
        {289998}
\bibfield{author}{\bibinfo{person}{S.G. Sedukhin}.} \bibinfo{year}{1994}\natexlab{}.
\newblock \showarticletitle{A new systolic architecture for pipeline prime factor {DFT}-algorithm}. In \bibinfo{booktitle}{\emph{Proceedings of 4th Great Lakes Symposium on VLSI}}. \bibinfo{publisher}{IEEE CS}, \bibinfo{address}{Notre Dame, IN, USA}, \bibinfo{pages}{40--45}.
\newblock
\urldef\tempurl%
\url{https://doi.org/10.1109/GLSV.1994.289998}
\showDOI{\tempurl}


\bibitem[Sedukhin(2012)]%
        {sedukhin_co-design_2012}
\bibfield{author}{\bibinfo{person}{Stanislav Sedukhin}.} \bibinfo{year}{2012}\natexlab{}.
\newblock \bibinfo{booktitle}{\emph{Co-design of {Extremely} {Scalable} {Algorithms}/{Architecture} for 3-{Dimensional} {Linear} {Transforms}}}.
\newblock \bibinfo{type}{{T}echnical {R}eport} Tech. Report 2012-001. \bibinfo{institution}{University of Aizu}, \bibinfo{address}{Japan}.
\newblock
\urldef\tempurl%
\url{https://u-aizu.ac.jp/files/page/research/techreport/2012-001.pdf}
\showURL{%
\tempurl}


\bibitem[Sedukhin et~al\mbox{.}(2009)]%
        {sedukhin_patent_2009}
\bibfield{author}{\bibinfo{person}{Stanislav. Sedukhin}, \bibinfo{person}{Toshiaki. Miyazaki}, {and} \bibinfo{person}{Kenichi. Kuroda}.} \bibinfo{year}{2009}\natexlab{}.
\newblock \bibinfo{title}{{Array Processor}}.
\newblock \bibinfo{howpublished}{\url{https://patents.google.com/patent/JP2009289256A/}}.
\newblock
\newblock
\shownote{Patent JP 2009289256A}.


\bibitem[Sedukhin et~al\mbox{.}(2015)]%
        {sedukhin_3d_2015}
\bibfield{author}{\bibinfo{person}{Stanislav Sedukhin}, \bibinfo{person}{Tomoya Sakai}, {and} \bibinfo{person}{Naohito Nakasato}.} \bibinfo{year}{2015}\natexlab{}.
\newblock \showarticletitle{{3D} {Discrete} {Transforms} with {Cubical} {Data} {Decomposition} on the {IBM} {Blue} {Gene}/{Q}}. In \bibinfo{booktitle}{\emph{Proc. of the 30th {International} {Conference} on {Computers} and {Their} {Applications}}}. \bibinfo{publisher}{Curran Associates, Inc.}, \bibinfo{address}{Honolulu, HW, USA}, \bibinfo{pages}{193--200}.
\newblock


\bibitem[Sedukhin et~al\mbox{.}(2022)]%
        {sedukhin_search_2022}
\bibfield{author}{\bibinfo{person}{Stanislav Sedukhin}, \bibinfo{person}{Yoichi Tomioka}, {and} \bibinfo{person}{Kohei Yamamoto}.} \bibinfo{year}{2022}\natexlab{}.
\newblock \showarticletitle{In {Search} of the {Performance}- and {Energy}-{Efficient} {CNN} {Accelerators}}.
\newblock \bibinfo{journal}{\emph{IEICE Transactions on Electronics}} \bibinfo{volume}{E105.C}, \bibinfo{number}{6} (\bibinfo{date}{June} \bibinfo{year}{2022}), \bibinfo{pages}{209--221}.
\newblock
\showISSN{0916-8524, 1745-1353}
\urldef\tempurl%
\url{https://doi.org/10.1587/transele.2021LHP0003}
\showDOI{\tempurl}


\bibitem[Sedukhin and Sedukhin(1994)]%
        {10.1007/3-540-58430-7_16}
\bibfield{author}{\bibinfo{person}{S.~G. Sedukhin} {and} \bibinfo{person}{I.~S. Sedukhin}.} \bibinfo{year}{1994}\natexlab{}.
\newblock \showarticletitle{Systematic approach and software tool for systolic design}. In \bibinfo{booktitle}{\emph{Parallel Processing: CONPAR 94 --- VAPP VI}}, \bibfield{editor}{\bibinfo{person}{Bruno Buchberger} {and} \bibinfo{person}{Jens Volkert}} (Eds.). \bibinfo{publisher}{Springer Berlin Heidelberg}, \bibinfo{address}{Berlin, Heidelberg}, \bibinfo{pages}{172--183}.
\newblock
\showISBNx{978-3-540-48789-0}
\urldef\tempurl%
\url{https://doi.org/10.1007/3-540-58430-7_16}
\showDOI{\tempurl}


\bibitem[Sedukhin et~al\mbox{.}(2010)]%
        {Sedukhin2010}
\bibfield{author}{\bibinfo{person}{Stanislav~G. Sedukhin}, \bibinfo{person}{Ahmed~S. Zekri}, {and} \bibinfo{person}{Toshiaki Myiazaki}.} \bibinfo{year}{2010}\natexlab{}.
\newblock \showarticletitle{{Orbital Algorithms and Unified Array Processor for Computing 2D Separable Transforms}}. In \bibinfo{booktitle}{\emph{2010 39th International Conference on Parallel Processing Workshops}}. \bibinfo{publisher}{IEEE}, \bibinfo{address}{San Diego, CA, USA}, \bibinfo{pages}{127--134}.
\newblock
\urldef\tempurl%
\url{https://doi.org/10.1109/ICPPW.2010.29}
\showDOI{\tempurl}


\bibitem[Sidiropoulos et~al\mbox{.}(2017)]%
        {7891546}
\bibfield{author}{\bibinfo{person}{Nicholas~D. Sidiropoulos}, \bibinfo{person}{Lieven De~Lathauwer}, \bibinfo{person}{Xiao Fu}, \bibinfo{person}{Kejun Huang}, \bibinfo{person}{Evangelos~E. Papalexakis}, {and} \bibinfo{person}{Christos Faloutsos}.} \bibinfo{year}{2017}\natexlab{}.
\newblock \showarticletitle{Tensor Decomposition for Signal Processing and Machine Learning}.
\newblock \bibinfo{journal}{\emph{IEEE Transactions on Signal Processing}} \bibinfo{volume}{65}, \bibinfo{number}{13} (\bibinfo{year}{2017}), \bibinfo{pages}{3551--3582}.
\newblock
\urldef\tempurl%
\url{https://doi.org/10.1109/TSP.2017.2690524}
\showDOI{\tempurl}


\bibitem[Smith and Torng(1985)]%
        {6158974}
\bibfield{author}{\bibinfo{person}{S.~P. Smith} {and} \bibinfo{person}{H.~C. Torng}.} \bibinfo{year}{1985}\natexlab{}.
\newblock \showarticletitle{Design of a fast inner product processor}. In \bibinfo{booktitle}{\emph{1985 IEEE 7th Symposium on Computer Arithmetic ({ARITH})}}. \bibinfo{publisher}{IEEE}, \bibinfo{address}{Urbana, Illinois, USA}, \bibinfo{pages}{38--43}.
\newblock
\urldef\tempurl%
\url{https://doi.org/10.1109/ARITH.1985.6158974}
\showDOI{\tempurl}


\bibitem[So et~al\mbox{.}(2022)]%
        {so2022primersearchingefficienttransformers}
\bibfield{author}{\bibinfo{person}{David~R. So}, \bibinfo{person}{Wojciech Mańke}, \bibinfo{person}{Hanxiao Liu}, \bibinfo{person}{Zihang Dai}, \bibinfo{person}{Noam Shazeer}, {and} \bibinfo{person}{Quoc~V. Le}.} \bibinfo{year}{2022}\natexlab{}.
\newblock \bibinfo{title}{Primer: Searching for Efficient Transformers for Language Modeling}.
\newblock
\newblock
\showeprint[arxiv]{2109.08668}~[cs.LG]
\urldef\tempurl%
\url{https://arxiv.org/abs/2109.08668}
\showURL{%
\tempurl}


\bibitem[Swartzlander et~al\mbox{.}(1978)]%
        {1674948}
\bibfield{author}{\bibinfo{person}{E.E. Swartzlander}, \bibinfo{person}{B.K. Gilbert}, {and} \bibinfo{person}{I.S. Reed}.} \bibinfo{year}{1978}\natexlab{}.
\newblock \showarticletitle{Inner Product Computers}.
\newblock \bibinfo{journal}{\emph{IEEE Trans. Comput.}} \bibinfo{volume}{C-27}, \bibinfo{number}{1} (\bibinfo{year}{1978}), \bibinfo{pages}{21--31}.
\newblock
\urldef\tempurl%
\url{https://doi.org/10.1109/TC.1978.1674948}
\showDOI{\tempurl}


\bibitem[Swartzlander~Jr(1989)]%
        {swartzlander1989wafer}
\bibfield{author}{\bibinfo{person}{Earl~E Swartzlander~Jr}.} \bibinfo{year}{1989}\natexlab{}.
\newblock \bibinfo{booktitle}{\emph{Wafer Scale Integration}}.
\newblock \bibinfo{publisher}{Kluwer Academic Publishers}, \bibinfo{address}{Boston, Dordrecht, London}. 523 pages.
\newblock


\bibitem[Thangarasa et~al\mbox{.}(2024)]%
        {thangarasa2024sparseiftsparseisofloptransformations}
\bibfield{author}{\bibinfo{person}{Vithursan Thangarasa}, \bibinfo{person}{Shreyas Saxena}, \bibinfo{person}{Abhay Gupta}, {and} \bibinfo{person}{Sean Lie}.} \bibinfo{year}{2024}\natexlab{}.
\newblock \bibinfo{title}{Sparse-IFT: Sparse Iso-FLOP Transformations for Maximizing Training Efficiency}.
\newblock
\newblock
\showeprint[arxiv]{2303.11525}~[cs.LG]
\urldef\tempurl%
\url{https://arxiv.org/abs/2303.11525}
\showURL{%
\tempurl}


\bibitem[Van De~Geijn and Watts(1997)]%
        {van_de_geijn_summa:_1997}
\bibfield{author}{\bibinfo{person}{R.~A. Van De~Geijn} {and} \bibinfo{person}{J. Watts}.} \bibinfo{year}{1997}\natexlab{}.
\newblock \showarticletitle{{SUMMA}: scalable universal matrix multiplication algorithm}.
\newblock \bibinfo{journal}{\emph{Concurrency: Practice and Experience}} \bibinfo{volume}{9}, \bibinfo{number}{4} (\bibinfo{date}{April} \bibinfo{year}{1997}), \bibinfo{pages}{255--274}.
\newblock
\showISSN{1040-3108, 1096-9128}


\bibitem[{Wikipedia contributors}(2025)]%
        {enwiki:1281191096}
\bibfield{author}{\bibinfo{person}{{Wikipedia contributors}}.} \bibinfo{year}{2025}\natexlab{}.
\newblock \bibinfo{title}{Advanced Matrix Extensions --- {Wikipedia}{,} The Free Encyclopedia}.
\newblock \bibinfo{howpublished}{\url{https://en.wikipedia.org/w/index.php?title=Advanced_Matrix_Extensions&oldid=1281191096}}.
\newblock
\newblock
\shownote{[Online; accessed 3-May-2025]}.


\bibitem[Wu et~al\mbox{.}(2023)]%
        {Wu_2023}
\bibfield{author}{\bibinfo{person}{Yannan~Nellie Wu}, \bibinfo{person}{Po-An Tsai}, \bibinfo{person}{Saurav Muralidharan}, \bibinfo{person}{Angshuman Parashar}, \bibinfo{person}{Vivienne Sze}, {and} \bibinfo{person}{Joel Emer}.} \bibinfo{year}{2023}\natexlab{}.
\newblock \showarticletitle{HighLight: Efficient and Flexible DNN Acceleration with Hierarchical Structured Sparsity}. In \bibinfo{booktitle}{\emph{56th Annual IEEE/ACM International Symposium on Microarchitecture}} \emph{(\bibinfo{series}{MICRO ’23})}. \bibinfo{publisher}{ACM}, \bibinfo{address}{Toronto, ON, Canada}, \bibinfo{pages}{1106–1120}.
\newblock
\urldef\tempurl%
\url{https://doi.org/10.1145/3613424.3623786}
\showDOI{\tempurl}


\bibitem[Xue et~al\mbox{.}(2023)]%
        {Xue_2023}
\bibfield{author}{\bibinfo{person}{Zi~Yu Xue}, \bibinfo{person}{Yannan~Nellie Wu}, \bibinfo{person}{Joel~S. Emer}, {and} \bibinfo{person}{Vivienne Sze}.} \bibinfo{year}{2023}\natexlab{}.
\newblock \showarticletitle{Tailors: Accelerating Sparse Tensor Algebra by Overbooking Buffer Capacity}. In \bibinfo{booktitle}{\emph{56th Annual IEEE/ACM International Symposium on Microarchitecture}} \emph{(\bibinfo{series}{MICRO ’23})}. \bibinfo{publisher}{ACM}, \bibinfo{address}{Toronto, ON, Canada}, \bibinfo{pages}{1347–1363}.
\newblock
\urldef\tempurl%
\url{https://doi.org/10.1145/3613424.3623793}
\showDOI{\tempurl}


\bibitem[Yen et~al\mbox{.}(2022)]%
        {yen2022s4highsparsityhighperformanceai}
\bibfield{author}{\bibinfo{person}{Ian En-Hsu Yen}, \bibinfo{person}{Zhibin Xiao}, {and} \bibinfo{person}{Dongkuan Xu}.} \bibinfo{year}{2022}\natexlab{}.
\newblock \bibinfo{title}{S4: a High-sparsity, High-performance AI Accelerator}.
\newblock
\newblock
\showeprint[arxiv]{2207.08006}~[cs.AR]
\urldef\tempurl%
\url{https://arxiv.org/abs/2207.08006}
\showURL{%
\tempurl}


\bibitem[Zhang et~al\mbox{.}(2021)]%
        {10.1145/3445814.3446702}
\bibfield{author}{\bibinfo{person}{Guowei Zhang}, \bibinfo{person}{Nithya Attaluri}, \bibinfo{person}{Joel~S. Emer}, {and} \bibinfo{person}{Daniel Sanchez}.} \bibinfo{year}{2021}\natexlab{}.
\newblock \showarticletitle{Gamma: leveraging Gustavson’s algorithm to accelerate sparse matrix multiplication}. In \bibinfo{booktitle}{\emph{Proceedings of the 26th ACM International Conference on Architectural Support for Programming Languages and Operating Systems}} (Virtual, USA) \emph{(\bibinfo{series}{ASPLOS '21})}. \bibinfo{publisher}{Association for Computing Machinery}, \bibinfo{address}{New York, NY, USA}, \bibinfo{pages}{687–701}.
\newblock
\showISBNx{9781450383172}
\urldef\tempurl%
\url{https://doi.org/10.1145/3445814.3446702}
\showDOI{\tempurl}


\end{thebibliography}

\end{document}